\documentclass[a4paper,onecolumn,11pt,accepted=2022-09-27
]{quantumarticle}
\pdfoutput=1
\usepackage[utf8]{inputenc}
\usepackage[english]{babel}
\usepackage[T1]{fontenc}
\usepackage{amsmath}
\usepackage{ascmac}
\usepackage{amsthm}
\usepackage{amssymb}
\usepackage{physics}
\usepackage{hyperref}
\usepackage{bm}
\usepackage{tikz}
\usepackage{lipsum}
\usepackage{float}
\newtheorem{theorem}{Theorem}

\begin{document}

\title{Role of matter coherence in entanglement due to gravity}

\author{Akira Matsumura}
\affiliation{Department of Physics, Kyushu University, Fukuoka, 819-0395, Japan}
\orcid{0000-0003-2634-6950}
\email{matsumura.akira@phys.kyushu-u.ac.jp}
\maketitle

\begin{abstract}
We investigate the quantum nature of gravity 
in terms of the coherence of quantum objects. 
As a basic setting, we consider two gravitating objects each in a superposition state of two paths.
The evolution of objects is described by the completely positive and trace-preserving (CPTP) map with a population-preserving property. 
This property reflects that the probability of objects being on each path is preserved.
We use the $\ell_1$-norm of coherence to quantify the coherence of objects.
In the present paper, the quantum nature of gravity
is characterized by an entangling map, which is a CPTP map with the capacity to create entanglement.
We introduce the entangling-map witness as an observable to test whether a given map is entangling. 
We show that, whenever the gravitating objects initially have a finite amount of the $\ell_1$-norm of coherence, the witness tests the entangling map due to gravity. 
Interestingly, we find that the witness can test such a quantum nature of gravity, even when the objects do not get entangled. 
This means that the coherence of gravitating objects always becomes the source of the entangling map due to gravity. 
We further discuss a decoherence effect and an experimental perspective in the present approach. 
\end{abstract}

\section{Introduction}

General relativity describes a gravitational field as a spacetime curvature, and quantum mechanics requires the quantum superposition principle.
The unified theory, quantum gravity, seems to predict the quantum superposition of spacetime curvature.
A lot of work is in great progress to prove such a quantum nature of gravity. 
The target in those works is the entanglement which comes from the quantum superposition of the gravitational field of a superposed massive object.
The entanglement due to gravity has been becoming a benchmark for the quantum nature of gravity \cite{Bose2017,Marletto2017}.
In sympathy with its importance, there were several theoretical works on various experimental proposals, for example, matter-wave interferometers \cite{Nguyen2020, Chevalier2020, vandeKamp2020, Miki2021, Tilly2021}, mechanical oscillators \cite{Krisnanda2020, Qvafort2020},
optomechanical systems 
\cite{Balushi2018,Miao2020,Matsumura2020,Miki2022}, and those hybrid systems \cite{Carney2021a,Pedernales2022, Matsumura2022a}. 
The detection of entanglement due to gravity also allows us to test classical models of gravity, such as the Newton-Schr\"{o}dinger approach \cite{Bahrami2014} and the classical channel gravity \cite{Kafri2014}.

In this paper, we deepen the understanding of the relation between the superposition of a massive object and the entanglement due to gravity. 
As a basic setting, we consider two massive objects each in a superposition state of paths. 
We introduce the $\ell_1$-norm measure to quantify the coherence of each superposed object \cite{Baumgratz2014}. 
It is demonstrated that the entanglement between the gravitating objects does not occur when their coherence is initially small.
We then examine the quantum evolution due to gravity. 
The evolution of the gravitating objects is represented by a completely positive and trace-preserving (CPTP) map with a population-preserving property. 
This property represents the fact that the probability of objects being on each path is preserved.
We explore when the evolution of the objects is given by an entangling map \cite{Harrow2003, Brandao2010}, which is defined as a CPTP map \cite{Nielsen2002} with the capacity to create entanglement. 
By the convex property of population-preserving non-entangling maps, we can introduce an entangling-map witness (a Hermitian operator) to test whether a given population-preserving CPTP map is entangling or not. 
Applying the theorems obtained in \cite{Matsumura2022b}, we derive the entangling-map witness to test whether the gravitational interaction between the objects leads to an entangling map. 
Through the above analysis, we find that the entangling map due to gravity is realized as long as each object initially has a finite amount of the $\ell_1$-norm of coherence. 
Interestingly, we observe that 
the entangling map due to gravity is realized even when the entanglement between the gravitating objects does not occur. 
The present result means that the coherence of massive objects always induces the entangling map due to gravity. 
We further discuss a decoherence effect and an experimental perspective in the present approach for testing the quantum nature of gravity. 

This paper is structured as follows. 
In Sec.\ref{sec:2}, we present the entanglement between two gravitating objects with coherence. 
It is exemplified that the entanglement does not occur when the coherence is small. 
In Sec.\ref{sec:3}, we introduce the formalism to examine an entangling map and its witness. 
In Sec.\ref{sec:4}, we find the witness for testing the entangling map due to gravity, which is given by the coherence of the objects. 
In Sec.\ref{sec:5}, we investigate a decoherence effect on the entangling map due to gravity, and we discuss an experimental advantage for verifying the entangling map. 
In Sec.\ref{sec:6}, we summarize this paper. 

\section{Entanglement due to gravity in quantum objects with coherence}
\label{sec:2}

\subsection{Time evolution of two massive objects each in a superposition}

We consider the time evolution of two massive objects interacting with each other through the Newtonian potential. The total Hamiltonian is 
\begin{equation}
\hat{H}=\hat{H}_\text{A}+\hat{H}_\text{B}+\hat{V}, 
\quad 
\hat{V}=-\frac{Gm_\text{A} m_\text{B}}{|\hat{\bm{x}}_\text{A}-\hat{\bm{x}}_\text{B}|},
\label{eq:H}
\end{equation}
where 
$\hat{H}_\text{A}$ and 
$\hat{H}_\text{B}$ are the Hamiltonians of objects A and B with masses
$m_\text{A}$ and 
$m_\text{B}$. 
The interaction Hamiltonian 
$\hat{V}$ is the quantized Newtonian potential, which is derived from the scattering amplitude in a linearized theory of quantized gravity around a Minkowski background, see, for example, \cite{Bose2022, Marshman2020}. 
The fact that the above Newtonian potential is not classical but an operator-valued function leads to the entanglement between two objects A and B. 
We assume that each object is superposed in two paths and that the paths of each object are determined by the Hamiltonians 
$\hat{H}_\text{A}$ and 
$\hat{H}_\text{B}$. 
Fig.\ref{fig:conf} in Sec.\ref{sec:2}.C presents an example of the configuration of two objects. 
Those objects acquire the phase shifts induced by the gravitational potential between them. 
We consider the initial state of the objects as 
\begin{equation}
|\Psi_\text{in} \rangle = \frac{1}{2} \sum_{a,b=\text{L},\text{R}}   |a\,; \text{in} \rangle_\text{A} \,  |b\,; \text{in} \rangle_\text{B},
\label{eq:Psii}
\end{equation}
where 
$|a\,;\text{in} \rangle_\text{A} \, (|b\,;\text{in} \rangle_\text{B})$ 
is the state of the particle A (B) with a localized wave packet with a peak at the position $a \, (b)$.
The state satisfies  
${}_\text{A} \langle a'\,;\text{in}|a\,;\text{in}\rangle_\text{A} \approx \delta_{a'a}\, ({}_\text{B} \langle b'\,;\text{in}|b\,;\text{in}\rangle_\text{B} \approx \delta_{b'b})$. 
The evolved state of objects A and B is given as  
\begin{align}
|\Psi_\text{out} \rangle 
&= e^{-i\hat{H}(t_\text{out}-t_\text{in})/\hbar}|\Psi_\text{in}\rangle
\nonumber 
\\
&=e^{-i(\hat{H}_\text{A}+\hat{H}_\text{B})(t_\text{out}-t_\text{in})/\hbar} 
\text{T}\exp \Big[ -\frac{i}{\hbar}\int^{t_\text{out}}_{t_\text{in}} dt \hat{V}_\text{I} (t)\Big] |\Psi_\text{in} \rangle \nonumber\\
&=
e^{-i(\hat{H}_\text{A}+\hat{H}_\text{B})(t_\text{out}-t_\text{in})/\hbar}
\text{T}
\exp \Big[-\frac{i}{\hbar}\int^{t_\text{out}}_{t_\text{in}} dt \hat{V}_\text{I} (t) \Big]
\frac{1}{2} \sum_{a,b=\text{L},\text{R}}|a\,;\text{in} \rangle_\text{A} \, |b\,;\text{in} \rangle_\text{B}
\nonumber 
\\
&\approx \frac{1}{2}
e^{-i(\hat{H}_\text{A}+\hat{H}_\text{B})(t_\text{out}-t_\text{in})/\hbar} 
\sum_{a,b=\text{L},\text{R}}e^{i\Phi_{ab}}\, |a\,;\text{in} \rangle_\text{A} \,  |b\,;\text{in} \rangle_\text{B}
\nonumber 
\\
&=
\frac{1}{2}
\sum_{a,b=\text{L},\text{R}}e^{i\Phi_{ab}}\,  |a\,;\text{out} \rangle_\text{A} \, |b\,;\text{out} \rangle_\text{B}, 
\label{eq:Sol}
\end{align}
where 
$\hat{V}_\text{I}(t)$ is the interaction Hamiltonian in the interaction picture with respect to 
$\hat{H}_\text{A}+\hat{H}_\text{B}$, and 
$|a\,;\text{out} \rangle_\text{A} = e^{-i\hat{H}_\text{A}(t_\text{out}-t_\text{in})/\hbar}|a\,;\text{in} \rangle_\text{A}$ and 
$|b\,;\text{out} \rangle_\text{B} = e^{-i\hat{H}_\text{B}(t_\text{out}-t_\text{in})/\hbar}|b\,;\text{in} \rangle_\text{B}$ are the states of wave packets of A and B with peaks at positions 
$a$ and 
$b$, respectively.
For the approximation in the fourth line, each object is assumed to be localized on each path during the evolution. 
The phase shift 
$\Phi_{ab}$ is evaluated along paths as 
\begin{equation}
\Phi_{ab}=\int^{t_\text{out}}_{t_\text{in}} \frac{dt}{\hbar} \frac{Gm_\text{A} m_\text{B}}{|\bm{x}^a_\text{A}(t)-\bm{x}^b_\text{B}(t)|},
\label{eq:PhiPQ}
\end{equation}
where 
$\bm{x}^a_\text{A}(t)$ and 
$\bm{x}^b_\text{B}(t)$ represent the trajectories of objects A and B, respectively. 
Using the initial and final density operators, 
$\rho_\text{out}=|\Psi_\text{out} \rangle \langle \Psi_\text{out}|$ and
$\rho_\text{in}=|\Psi_\text{in} \rangle \langle \Psi_\text{in}|$, 
we can rewrite the above evolution as 
\begin{widetext}
\begin{align}
\rho_\text{out}
&=|\Psi_\text{out} \rangle \langle \Psi_\text{out}|
\nonumber 
\\
&=
\frac{1}{4}
\sum_{a,b=\text{L},\text{R}}
\sum_{a',b'=\text{L},\text{R}}e^{i(\Phi_{ab}-\Phi_{a'b'})}\,  |a\,;\text{out} \rangle_\text{A} \langle a'\,;\text{out} | \otimes  |b\,;\text{out} \rangle_\text{B} \langle b'\,;\text{out}|
\nonumber 
\\
&=
\
\sum_{a,b=\text{L},\text{R}}
\sum_{a',b'=\text{L},\text{R}} (\mathcal{E}_\text{G})_{aba'b'} \,  \hat{M}_a \otimes \hat{N}_b \, \rho_\text{in} \,\hat{M}^\dagger_{a'} \otimes \hat{N}^\dagger_{b'},
\label{eq:Sol2}
\end{align}
\end{widetext}
where 
$(\mathcal{E}_\text{G})_{aba'b'}=e^{i(\Phi_{ab}-\Phi_{a'b'})}$,
$\hat{M}_a=|a\,;\text{out} \rangle_\text{A} \langle a\,;\text{in}|$ and 
$\hat{N}_b=|b\,;\text{out} \rangle_\text{A} \langle b\,;\text{in}|$. 
A general feature of this evolution will be discussed in Sec.\ref{sec:3} based on quantum information theory. 

\subsection{Coherence of the initial state of objects}

We consider the initial state of two massive objects at 
$t=t_\text{in}$ as
\begin{equation}
\rho_\text{in} =\rho_\text{A} \otimes \rho_\text{B},
\label{eq:rhoi}
\end{equation}
where 
\begin{align}
\rho_\text{A}
&=\frac{1}{2} 
\big(
|\text{L}\,;\text{in} \rangle_\text{A} \langle \text{L}\,;\text{in} |
+|\text{R}\,;\text{in} \rangle_\text{A} \langle \text{R}\,;\text{in} |
+c_1|\text{L}\,;\text{in} \rangle_\text{A} \langle \text{R}\,;\text{in} |
+c_1|\text{R}\,;\text{in} \rangle_\text{A} \langle \text{L}\,;\text{in} |
\big),
\label{eq:rhoA}
\\
\rho_\text{B}
&=\frac{1}{2} 
\big(
|\text{L}\,;\text{in} \rangle_\text{B} \langle \text{L}\,;\text{in}|
+|\text{R}\,;\text{in} \rangle_\text{B} \langle \text{R}\,;\text{in}|
+c_2|\text{L}\,;\text{in}\rangle_\text{B} \langle \text{R}\,;\text{in}|
+c_2|\text{R}\,;\text{in} \rangle_\text{B} \langle \text{L}\,;\text{in} |
\big).
\label{eq:rhoB}
\end{align}
The parameters 
$c_1$ and 
$c_2$ 
($-1 \leq c_1 \leq 1, \, -1 \leq c_2 \leq 1$) characterize the interference effects of each object. 
When 
$c_1=\pm 1$, the initial state of object A is 
$(|\text{L}\,;\text{in} \rangle_\text{A} \pm  |\text{R}\,;\text{in} \rangle_\text{A})/\sqrt{2}$, which is the superposition state of two wave packets around the positions 
L and R. 
When 
$c_1=0$, the initial state is 
$(|\text{L}\,;\text{in} \rangle_\text{A} \langle \text{L}\,;\text{in}| + |\text{R}\,;\text{in} \rangle_\text{A} \langle \text{R}\,;\text{in}|)/2$, and object A has no interference effect. 
The same feature is observed in the initial state of object B for 
$c_2=\pm 1$ or 
$c_2=0$.
The degree of the interference effect (the quantum superposition) is quantified by the $\ell_1$-norm \cite{Baumgratz2014} of coherence,
\begin{equation}
C(\rho)=\sum_{i\neq j} |\langle i | \rho |j \rangle|, 
\label{eq:C}
\end{equation}
where 
$\{ |i \rangle \}_i$ is a complete orthonormal basis. 
The $\ell_1$-norm of coherence for the initial state of each object are
\begin{equation}
C(\rho_\text{A})=|c_1|, \quad C(\rho_\text{B})=|c_2|, 
\label{eq:CACB}
\end{equation}
where the bases 
$\{|\text{L}\,;\text{in} \rangle_\text{A},|\text{R}\,;\text{in} \rangle_\text{A} \}$ and 
$\{|\text{L}\,;\text{in} \rangle_\text{B},|\text{R}\,;\text{in} \rangle_\text{B} \}$ were chosen. 

To understand what the parameters 
$c_1$ and 
$c_2$ characterize in a more realistic situation, we focus on the setting in \cite{Bose2017}, where the spatial superposition of an object with a spin is assumed to be created by a Stern-Gerlach apparatus. 
In such a setting, the states 
$|\text{L}\,;\text{in} \rangle_\text{X}$ and
$|\text{R}\,;\text{in} \rangle_\text{X}$ of object
$\text{X}=\text{A},\text{B}$ are regarded as 
$|\text{L}\,;\text{in} \rangle_\text{X}=|\text{C}\,;\text{in} \rangle_\text{X} |\!\!\uparrow\rangle_\text{X}$ and
$|\text{R}\,;\text{in} \rangle_\text{X}=|\text{C}\,;\text{in} \rangle_\text{X} |\!\!\downarrow\rangle_\text{X}$, respectively. 
Here, 
$|\text{C}\,;\text{in} \rangle_\text{X}$ is the state of the external degrees of freedom, and 
$|\!\!\uparrow\rangle_\text{X}$ and 
$|\!\!\downarrow\rangle_\text{X}$ are the states of the spin internal degrees of freedom. 
Each object with spin moves in an external magnetic field within the Stern-Gerlach apparatus. 
The initial states of objects A and B are 
\begin{align}
\rho_\text{A}
&=|\text{C}\,;\text{in} \rangle_\text{A} \langle \text{C}\,;\text{in}| 
\otimes \frac{1}{2} 
\big(
|\!\!\uparrow\rangle_\text{A} \langle \uparrow\!\!|
+|\!\!\downarrow\rangle_\text{A} \langle \downarrow\!\!|
+c_1|\!\!\uparrow\rangle_\text{A} \langle\downarrow\!\!|
+c_1|\!\!\downarrow\rangle_\text{A} \langle \uparrow\!\!|
\big),
\label{eq:rhoA2}
\\
\rho_\text{B}
&=|\text{C}\,;\text{in} \rangle_\text{B} \langle \text{C}\,;\text{in}| 
\otimes \frac{1}{2} 
\big(
|\!\!\uparrow\rangle_\text{B} \langle \uparrow\!\!|
+|\!\!\downarrow\rangle_\text{B} \langle \downarrow\!\!|
+c_2|\!\!\uparrow\rangle_\text{B} \langle\downarrow\!\!|
+c_2|\!\!\downarrow\rangle_\text{B} \langle \uparrow\!\!|
\big).
\label{eq:rhoB2}
\end{align}
If the parameters 
$c_1$ and 
$c_2$ are nonzero, each object is spatially superposed in the apparatus and has an interference effect. 
From a practical viewpoint, the initial state of each spin is prepared through an experimental procedure. 
When 
$c_1=c_2=0$, each spin state is completely mixed, and then the apparatus cannot create the spatial superposition of objects. 
Roughly speaking, the parameters
$c_1$ and 
$c_2$ determine whether a spin state with coherence is prepared or not. 

\subsection{Effect of coherence on entanglement due to gravity}

We introduce the notion of a separable (non-entangled) state and evaluate the entanglement between the gravitating objects considered in the previous subsection. 
A density operator 
$\rho_\text{AB}$ of a bipartite system AB is said to be separable \cite{Werner1989} if the density operator is written as
\begin{equation}
\rho_\text{AB}=\sum_{i} p_i \rho^i_\text{A} \otimes \rho^i_\text{B},
\label{eq:sep}
\end{equation}
where 
$p_i$ is a probability, and 
$\rho^i_\text{A}$ and 
$\rho^i_\text{B}$ are the density operators of each of the subsystems A and B. 
The positive partial
transpose criterion \cite{Peres1996, Horodecki1996} and the negativity \cite{Vidal2002} are useful to judge whether a given density operator 
$\rho_\text{AB}$ is separable or not. 
For a density operator 
$\rho_\text{AB}$ of a bipartite system AB, we define the partial transpose 
$\rho^{\text{T}_\text{A}}_\text{AB}$ with the components given by
\begin{equation}
{}_\text{A} \langle a | {}_\text{B} \langle b| \rho^{\text{T}_\text{A}}_\text{AB} |a' \rangle_\text{A} |b' \rangle_\text{B}= {}_\text{A} \langle a' | {}_\text{B} \langle b| \rho_\text{AB} |a \rangle_\text{A} |b' \rangle_\text{B}
\end{equation}
for a basis 
$\{ |a \rangle_\text{A} |b \rangle_\text{B} \}_{a,b}$ of the Hilbert space $\mathcal{H}_\text{A} \otimes \mathcal{H}_\text{B}$ of the system AB.  
It is shown that if the density operator 
$\rho_\text{AB}$ is separable then its partial transposition 
$\rho^{\text{T}_\text{A}}_\text{AB}$ is positive semidefinite (has only non-negative eigenvalues). 
Hence, if the partial transposition 
$\rho^{\text{T}_\text{A}}_\text{AB}$ has a negative eigenvalue then the density operator $\rho_\text{AB}$ is entangled. 
This is called the positive partial transpose (PPT) criterion \cite{Peres1996, Horodecki1996}. 
In particular, the PPT criterion is the necessary and sufficient condition for a two-qubit system 
($\mathbb{C}^2 \otimes \mathbb{C}^2$) and a qubit-qutrit system 
($\mathbb{C}^2 \otimes \mathbb{C}^3$) \cite{Horodecki1996}. 
We obtain an entanglement measure (quantifier) based on the PPT criterion as 
\begin{equation}
\mathcal{N}=\sum_{\lambda_i<0} |\lambda_i|, 
\end{equation}
where 
$\lambda_i$ is the eigenvalue of the partial transposition $\rho^{\text{T}_\text{A}}$. 
This is called the negativity \cite{Vidal2002}. 
The PPT criterion means that a density operator 
$\rho_\text{AB}$ with nonzero negativity is entangled. 

\begin{figure}[htbp]
\centering
  \includegraphics[width=0.60\linewidth]{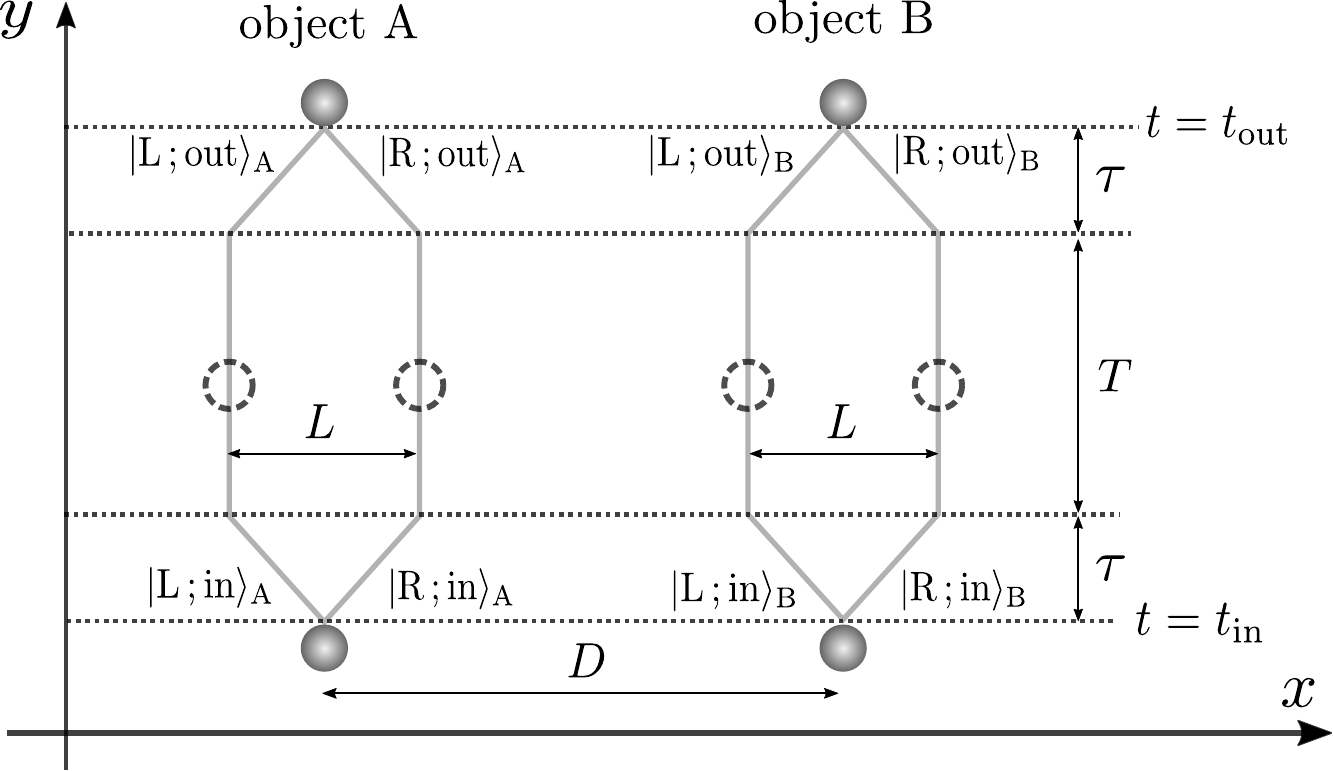}
  \caption{A configuration of two massive objects A and B. 
  In this setting, the length scale of superposition is 
  $L$ and the separation between two objects is 
  $D$. 
  The superposition is preserved during the time $T$. }
  \label{fig:conf}
\end{figure}

We demonstrate the entanglement due to gravity adopting the configuration shown in Fig. \ref{fig:conf}.  
Two massive objects are separated by a distance 
$D$ in the x-direction. 
Each object is in a spatial superposition along the x-direction, whose length scale is 
$L$.
The splitting and refocusing times of the trajectory of each object are 
$\tau$, and the superposition is kept during a time 
$T$. 
The concrete trajectories are
\begin{equation}
\bm{x}^a_\text{A}(t)=[x^a(t), v_y t,0]^\text{T},\quad
\bm{x}^a_\text{B}(t)=[x^a(t)+D, v_y t,0]^\text{T},
\end{equation}
where 
$v_y>0$ is the velocity in the y-direction and 
\begin{equation}
x^a(t)=\epsilon^a
\left \{
\begin{array}{lll}
 v_x t & t_\text{in}=0 \leq t \leq \tau \\
 v_x\tau &  \tau \leq t \leq T+\tau \\ 
 v_x (T+\tau-t)+v_x \tau & T+\tau \leq t \leq T+2\tau=t_\text{out}
\end{array}
\right.
\end{equation}
with 
$\epsilon^\text{L}=-\epsilon^\text{R}=-1$ and 
$v_x (>0)$ is the velocity in the x-direction. 
In this setting, for simplicity, we assumed that the two objects are instantaneously accelerated at $t=t_\text{in}, t=t_\text{in}+\tau$ and 
$t=t_\text{in}+\tau+T$. 
If each object has a spin degree of freedom, in this setting, a  non-uniform magnetic field is instantaneously injected at each time to split and refocus each trajectory of the objects.
The present setting is just used for a theoretical demonstration of entanglement due to gravity.
For a more realistic setting as in Ref.\cite{Bose2017}, the fluctuations of the magnetic field lead to a decoherence effect. 
The effect might be absorbed in the parameters $c_1,c_2$ or the damping (decoherence) rates $\gamma_\text{A}, \gamma_\text{B}$ discussed in Sec.\ref{sec:5}. 
In the present paper, we do not analyze the effects in detail. 

\begin{figure}[htbp]
  \centering
  \includegraphics[width=0.80\linewidth]{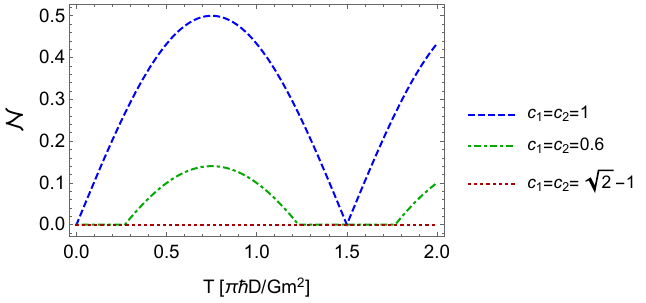}
  \caption{The behavior of the negativity (dimensionless quantity) of two gravitating objects with various degrees of coherence.}
  \label{fig:ngtvty}
\end{figure}

The phase shift induced by the gravitational potential is  
\begin{align}
\Phi_{ab}
&= \int^{t_\text{out}}_{t_\text{in}} \frac{dt}{\hbar} \frac{Gm_\text{A}m_\text{B}}{|\bm{x}^a_\text{A}(t) - \bm{x}^b_\text{B}(t)|}
\nonumber 
\\
&=\frac{Gm_\text{A}m_\text{B}}{\hbar}
\Big \{
\frac{4\tau}{(\epsilon_b-\epsilon_a)L }
\ln\Big[1+\frac{\epsilon_b-\epsilon_a}{2} \frac{L}{D}\Big]
+
\frac{T}{D+(\epsilon_b-\epsilon_a)L/2} 
\Big \}
\nonumber 
\\
&\approx 
\frac{Gm_\text{A}m_\text{B}}{D+(\epsilon_b-\epsilon_a)L/2} \frac{T}{\hbar} ,
\label{eq:PhiPQ2}
\end{align}
where 
$L=2v_x \tau$ and 
$T \gg \tau$ is assumed for the approximation in the last line. 
The final state 
$\rho_\text{out}$ is obtained from the initial state 
$\rho_\text{in}$ \eqref{eq:rhoi} and Eq.\eqref{eq:Sol2}. 
We can compute the negativity of 
$\rho_\text{out}$ as
\begin{align}
 \mathcal{N}&=\max[0,-\lambda], 
 \nonumber 
 \\
 \lambda&=\frac{1}{4}
 \Big[
 1-|c_1||c_2|
 -\sqrt{(|c_1|-|c_2|)^2+4|c_1||c_2|
 \sin^2
\Big(
\frac{\Phi_\text{RL}+\Phi_\text{LR}-\Phi_\text{LL}-\Phi_\text{RR}}{2}
\Big)}
 \Big],
 \label{eq:lambdamin}
\end{align}
where 
\begin{equation}
 \Phi_\text{LL}=\Phi_\text{RR}=\frac{Gm_\text{A}m_\text{B}}{D} \frac{T}{\hbar}, 
 \quad 
 \Phi_\text{LR}=\frac{Gm_\text{A}m_\text{B}}{D+L}\frac{T}{\hbar}, 
 \quad 
 \Phi_\text{RL}=\frac{Gm_\text{A}m_\text{B}}{D-L}\frac{T}{\hbar}.
 \label{eq:phis}
\end{equation}
Fig.\ref{fig:ngtvty} shows the negativity as a function of 
$T$ for the fixed masses $m_\text{A}=m_\text{B}=m$, the same degree of the coherence
$c_1=c_2$, and the fixed distance
$D=2L$. 
The blue dashed curve given for 
$c_1=c_2=1$ presents a large generation of the negativity.
On the other hand, the green dot-dashed curve shows a small amount of entanglement for 
$c_1=c_2=0.6$. 
The brown dotted curve 
given for 
$c_1=c_2=\sqrt{2}-1$
corresponds to the case without entanglement generation. 
To understand why such a critical value $\sqrt{2}-1$ emerges, an analysis in terms of quantum information might be needed. 
However, the entanglement degradation is explained as follows. 
The incoherent noise during the preparation of the initial state causes a small initial coherence.
The noise makes it difficult to measure the phase shifts and the entanglement generation due to gravity.
The green dot-dashed and brown dotted curves give us a lesson about the relation between the quantum superposition of objects and the entanglement due to gravity. 
For nonzero 
$c_1$ and 
$c_2$, each of objects A and B has a coherence, that is, each is spatially superposed to some extent. 
The objects with nonzero coherence may induce the quantum superposition of gravitational fields (or spacetime curvature). 
However, the entanglement due to gravity does not appear in those objects.

In the following, we will introduce the notion of an entangling map and clarify that the coherence of objects can be the source of the entangling map due to gravity.

\section{Path-entangling evolution and its witness}
\label{sec:3}

\subsection{Characterization of path-entangling evolution}

We begin with the definition of a completely positive trace-preserving (CPTP) map, a separable map, and a non-entangling map. 
A CPTP map is a general evolution of a quantum system, which can describe a unitary evolution and the dynamics of an open system. 
A CPTP map 
$\Phi$ on a density operator 
$\rho$ is given as 
\begin{equation}
\Phi[\rho]=
 \sum_\ell \hat{K}_\ell \, \rho \, \hat{K}^\dagger_\ell, 
\label{eq:Kraus}
\end{equation}
where 
$\hat{K}_\ell$ are called Kraus operators, which satisfy
$\sum_\ell \hat{K}^\dagger_\ell \hat{K}_\ell=\hat{\mathbb{I}}$ \cite{Nielsen2002}. 
The other maps, a separable map and a non-entangling map, form a subset of CPTP maps. 

A separable map \cite{Rains1997,Vedral1998} is a CPTP map with $\hat{K}_\ell=\hat{A}_\ell \otimes \hat{B}_\ell$, where 
$\hat{A}_\ell$ and 
$\hat{B}_\ell$ are local operators on two subsystems A and B, respectively. 
Any local operations and classical communication (LOCC), which characterizes the operational features of entanglement, is included in a class of separable maps \cite{Rains1997, Vedral1998, Horodecki2009}. 
In addition, it is known that the set of LOCC operations is a proper subset of separable maps \cite{Chitambar2014}.

As a larger class of separable maps, we can introduce a non-entangling map defined by a CPTP map that does not generate entanglement in any separable state \cite{Harrow2003, Brandao2010}. 
Namely, a CPTP map 
$\Phi$ such that 
$\Phi[\rho]$ is separable for any separable state $\rho$ is a non-entangling map. 
If an initial separable state 
$\rho$ evolves by a CPTP map 
$\Phi$
and the evolved state 
$\Phi[\rho]$ is entangled, then the CPTP map 
$\Phi$ is called an entangling map. 
In the previous section, we observed the entanglement between two gravitating objects which are initially in a separable state. 
Hence, the gravitational interaction leads to an entangling map on the objects' state. 

We next focus on the evolution of path-superposed objects as discussed in the previous section. 
The probability of such objects being on each path is preserved during evolution. 
This property is called population preserving in \cite{Carney2021a}, which is represented as 
\begin{widetext}
\begin{equation}
\text{Tr}\big[
 |a\,;\text{out} \rangle_\text{A} \langle a\,;\text{out} | \otimes   |b\,;\text{out} \rangle_\text{B} \langle b\,;\text{out} | \, \rho_\text{out} \big ]
 =\text{Tr} \big[
 |a\,;\text{in} \rangle_\text{A} \langle a\,;\text{in} | \otimes   |b\,;\text{in} \rangle_\text{B} \langle b\,;\text{in} | \, \rho_\text{in} \big],
 \label{eq:PP}
\end{equation}
\end{widetext}
where 
$|a\,;\text{in(out)} \rangle_\text{A}$ and 
$|b\,;\text{in(out)} \rangle_\text{B}$ are the states of A and B with wave packets around positions
$a$ and 
$b$
at a time 
$t_\text{in(out)}$, respectively.
$\rho_\text{in(out)}$ is the total density operator of objects A and B at $t_\text{in(out)}$. 
We assume that the evolution of the objects is given by a population-preserving CPTP map
$\Phi$. 
The representation of a population-preserving CPTP map
$\Phi$ is obtained in \cite{Matsumura2022b}:
\begin{theorem}
Let 
$\mathcal{H}^\text{in}_\text{A} \otimes \mathcal{H}^\text{in}_\text{B}$ and 
$\mathcal{H}^\text{out}_\text{A} \otimes \mathcal{H}^\text{out}_\text{B}$
be the Hilbert space with the complete bases 
$\{ |a\,;\text{in} \rangle_\text{A} |b\,;\text{in} \rangle_\text{B} \}_{a=1,2,\dots,d_\text{A},b=1,2,\dots,d_\text{B}}$ and 
$\{ |a\,;\text{out} \rangle_\text{A} |b\,;\text{out} \rangle_\text{B} \}_{a=1,2,\dots,d_\text{A},b=1,2,\dots,d_\text{B}}$, respectively. 
A map 
$\Phi :\rho_\text{in} \mapsto \rho_\text{out}=\Phi[\rho_\text{in}] $, where 
$\rho_\text{in}$ and 
$\rho_\text{out}$ are density operators each on  
$\mathcal{H}^\text{in}_\text{A} \otimes \mathcal{H}^\text{in}_\text{B}$ and
$\mathcal{H}^\text{out}_\text{A} \otimes \mathcal{H}^\text{out}_\text{B}$, 
is a population-preserving CPTP map  if and only if the map
$\Phi$ is represented by 
\begin{equation}
\Phi[\rho_\text{in}]
 =\sum^{d_\text{A}}_{a,a'=1} 
 \sum^{d_\text{B}}_{b,b'=1} \mathcal{E}_{aba'b'} 
 \hat{M}_a \otimes  \hat{N}_b \, \rho_\text{in}  \, 
 \hat{M}^\dagger_{a'} \otimes  \hat{N}^\dagger_{b'},
\label{eq:CPTPPP}
\end{equation}
where 
$\hat{M}_a=|a\,;\text{out} \rangle_\text{A} \langle a\,;\text{in}|$,  
$\hat{N}_b=|b\,;\text{out} \rangle_\text{B} \langle b\,;\text{in}|$ 
and 
the coefficients 
$\mathcal{E}_{aba'b'}$ form a 
$d_\text{A} d_\text{B} \times d_\text{A} d_\text{B}$ non-negative matrix
$\mathcal{E}$ with the diagonal elements 1 (
$\mathcal{E}_{abab}=1$). 
Here, a matrix 
$\mathcal{E}$ is said to be non-negative if 
$\bm{w}^\dagger \mathcal{E} \bm{w} \geq 0$ for all complex vectors 
$\bm{w} \in \mathbb{C}^{d_\text{A} d_\text{B}}$ or equivalently, 
$\mathcal{E}$ has only non-negative eigenvalues. 
\end{theorem}
Theorem 1 says that the properties of a population-preserving CPTP  map 
$\Phi$ are determined by 
$\mathcal{E}_{aba'b'}$.
The following theorems in \cite{Matsumura2022b} play an important role to judge whether a population-preserving CPTP  map 
$\Phi$ is entangling or not.
\begin{theorem}
Let 
$\Phi$ be the population-preserving CPTP  map in Theorem 1. 
$\Phi$ is a separable map if and only if the 
$d_\text{A} d_\text{B} \times d_\text{A} d_\text{B}$ matrix 
$\mathcal{E}$ with elements 
$\mathcal{E}_{aba'b'}$ in the representation of 
$\Phi$ is separable, that is, 
each element $\mathcal{E}_{aba'b'}$ has a following form,
\begin{equation}
\mathcal{E}_{aba'b'}
=\sum_{k} \lambda_k \, (\mathcal{A}_k)_{aa'} \, (\mathcal{B}_k)_{bb'},
\label{eq:sep2}
\end{equation}
where 
$\lambda_k \geq 0$, and 
$\mathcal{A}_k$ and 
$\mathcal{B}_k$ are non-negative matrices with components 
$(\mathcal{A}_k)_{aa'}$ and
$(\mathcal{B}_k)_{bb'}$, respectively.
\end{theorem}
\begin{theorem}
Let 
$\Phi$ be the population-preserving CPTP  map in Theorem 1.  
$\Phi$ is inseparable if and only if 
$\Phi$ is entangling.
\end{theorem}
Theorem 2 is derived from the theorem in \cite{Cirac2001}, in which the non-negative matrix 
$\mathcal{E}$ plays a similar role to the Choi matrix \cite{Jamiolkowski1972, Choi1975}. 
Theorem 3 follows from a straightforward calculation with the help of Theorem 2. 
These theorems mean that the inseparability of $\mathcal{E}$ associated with a population-preserving CPTP  map 
$\Phi$ determines whether 
$\Phi$ is entangling. 

\subsection{Entangling-map witness and its construction}

In this subsection, we introduce an observable (a Hermitian operator) to verify an entangling map, which will connect the entangling map due to gravity with the coherence of massive objects. 
For this purpose,
we focus on the following set of density operators,
\begin{equation}
S_\text{NE}[\rho_\text{in}] =\{ \, \sigma \, | \, \sigma=\Lambda[\rho_\text{in}], \Lambda \in \text{population-preserving non-entangling  maps} \, \},
\label{eq:S0}
\end{equation}
where 
$\rho_\text{in}$ is the initial state of two objects (for example, it is given by Eq. \eqref{eq:rhoi}). 
Using this set $S_\text{NE}[\rho_\text{in}]$, we can consider the possibility that the gravitational dynamics of a quantum object is described by a non-entangled map such as LOCC.
For example, in Ref.\cite{Kafri2014}, a LOCC model of gravity using continuous measurement and feedback was proposed.
In this case, even if gravitating objects are in a quantum superposition, those dynamics is described by a non-entangling map. 

\begin{figure}[htbp]
  \centering
  \includegraphics[width=0.70\linewidth]{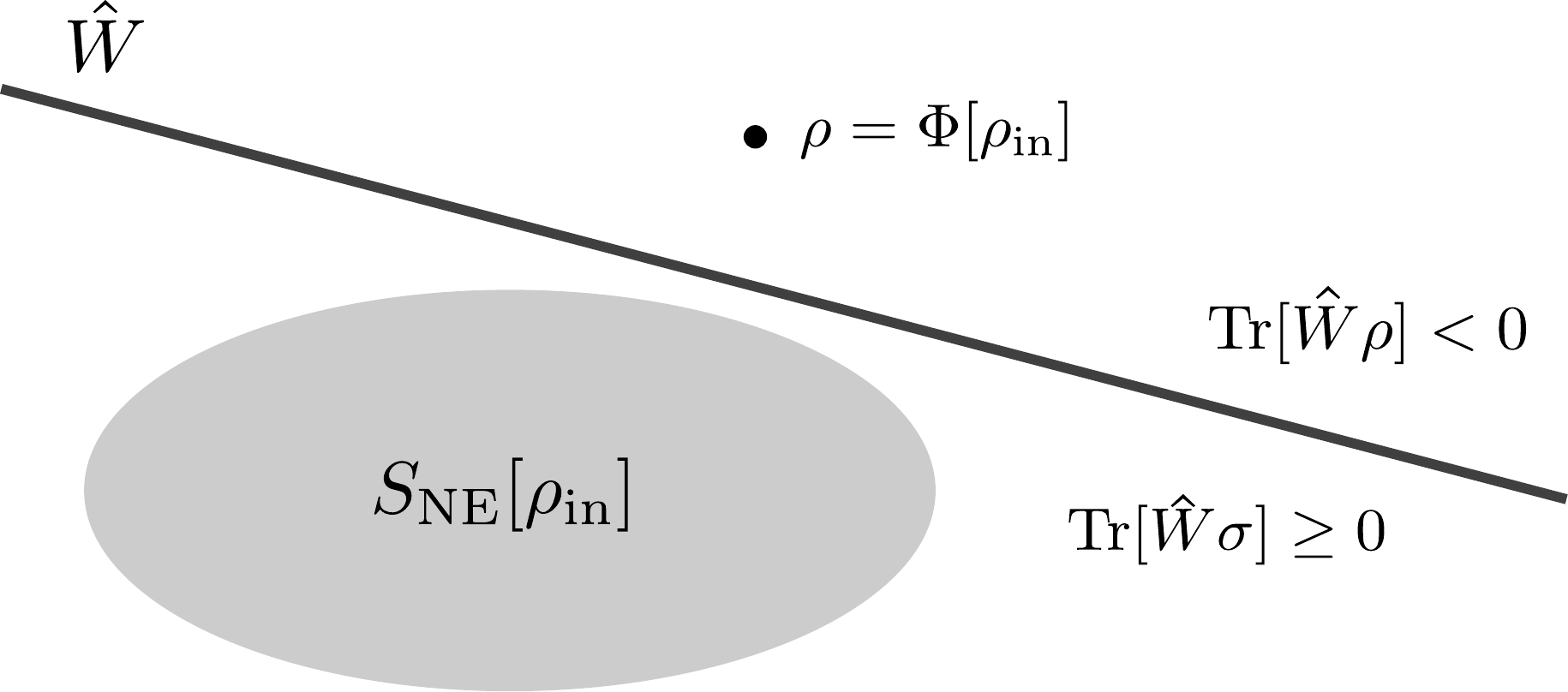}
  \caption{The hyperplane between the convex set $S_\text{NE}[\rho_\text{in}]$ and the point $\rho \notin S_\text{NE}[\rho_\text{in}]$ corresponds to the geometric meaning of the Hermitian operator $\hat{W}$. 
  Its positive or negative expectation value is assigned for the regions above or below the hyperplane.}
  \label{fig:W}
\end{figure}

We note that the following statement holds for all 
$0 \leq p \leq 1$:
\begin{align}
\Lambda_1, \Lambda_2 &\in \text{population-preserving non-entangling maps}
\nonumber 
\\
&\Rightarrow \, p \Lambda_1 +(1-p) \Lambda_2 \in \text{population-preserving non-entangling maps}. 
\end{align}
Hence the set
$S_\text{NE}[\rho_\text{in}]$ is convex, that is, for all 
$0\leq p \leq 1$,
\begin{equation}
\sigma_1, \sigma_2 \in S_\text{NE}[\rho_\text{in}] \,  \Rightarrow \, p \sigma_1 +(1-p) \sigma_2 \in S_\text{NE}[\rho_\text{in}] . 
\end{equation}
Since 
$S_\text{NE}[\rho_\text{in}]$ is convex, by the Hahn-Banach separation theorem (or similar discussions on entanglement witness \cite{Horodecki1996}), we always find a Hermitian operator 
$\hat{W}$ such that 
\begin{align}
\text{Tr}[\hat{W} \rho ] <0 \text{ for a given } \rho \notin S_\text{NE}[\rho_\text{in}]
\, \text{ and } \, \text{Tr}[\hat{W} \sigma] \geq0 \text{ for any } \sigma \in S_\text{NE}[\rho_\text{in}].
\label{eq:witness}
\end{align}
A geometrical interpretation of this fact is shown in Fig.\ref{fig:W}. 
The Hermitian operator 
$\hat{W}$ is the observable to test whether a given map is entangling or not. 
Let the density operator 
$\rho \notin S_\text{in} $ be given by 
$\rho=\Phi[\rho_\text{in}]$ with a population-preserving entangling map 
$\Phi$. 
Then, the signature of entangling map is confirmed from the negative expectation value 
$\text{Tr}[\hat{W}\rho] <0$.
Such an observable 
$\hat{W}$ is called the entangling-map witness in this paper.
We can obtain a strategy to find an entangling-map witness from the representation theorem of a population-preserving CPTP  map (Theorem 1). 
Using the representation of the map 
$\Phi$ given in \eqref{eq:CPTPPP}, we can write the expectation value of
$\hat{W}$ 
for the density operator 
$\rho=\Phi[\rho_\text{in}]$ as 
\begin{align}
\text{Tr}[\hat{W} \rho]
&=\text{Tr} \big[\hat{W} \Phi[\rho_\text{in}] \big]
\nonumber 
\\
&= \text{Tr}
\Big[\hat{W} \sum^{d_\text{A}}_{a,a'=1}\sum^{d_\text{B}}_{b,b'=1} \mathcal{E}_{aba'b'}\hat{M}_a \otimes  \hat{N}_b \, \rho_\text{in}  \, 
 \hat{M}^\dagger_{a'} \otimes  \hat{N}^\dagger_{b'}
 \Big]
\nonumber 
\\
&
=\sum^{d_\text{A}}_{a,a'=1}\sum^{d_\text{B}}_{b,b'=1} \mathcal{E}_{aba'b'} \mathcal{W}_{a'b'ab}
\nonumber 
\\
&
=\text{tr}[\mathcal{E} \, \mathcal{W}],
\label{eq:trEW}
\end{align}
where in the second line
$\mathcal{W}_{a'b'ab}$ was defined by 
\begin{equation}
\mathcal{W}_{a'b'ab}=\text{Tr}[\hat{W} \hat{M}_a \otimes  \hat{N}_b \, \rho_\text{in}  \, 
 \hat{M}^\dagger_{a'} \otimes  \hat{N}^\dagger_{b'}],
 \label{eq:w}
\end{equation}
and in the last line the $d_\text{A}d_\text{B} \times d_\text{A}d_\text{B}$ matrices 
$\mathcal{E}$ and 
$\mathcal{W}$ with elements 
$\mathcal{E}_{aba'b'}$ and 
$\mathcal{W}_{aba'b'}$ were introduced to rewrite the expectation value in a simple form. 
Similarly, we obtain the expectation value of 
$\hat{W}$ for the density operator 
$\sigma=\Lambda[\rho_\text{in}]$ as
\begin{align}
\text{Tr}[\hat{W} \sigma]
&=\text{Tr} \big[\hat{W} \Lambda[\rho_\text{in}] \big]
\nonumber 
\\
&= \text{Tr}
\Big[\hat{W} \sum^{d_\text{A}}_{a,a'=1}\sum^{d_\text{B}}_{b,b'=1} \mathcal{F}_{aba'b'}\hat{M}_a \otimes  \hat{N}_b \, \rho_\text{in}  \, 
 \hat{M}^\dagger_{a'} \otimes  \hat{N}^\dagger_{b'}
 \Big]
\nonumber 
\\
&
=\sum^{d_\text{A}}_{a,a'=1}\sum^{d_\text{B}}_{b,b'=1} \mathcal{F}_{aba'b'} \mathcal{W}_{a'b'ab}
\nonumber 
\\
&
=\text{tr}[\mathcal{F} \, \mathcal{W}].
\label{eq:trFW}
\end{align}
Here, 
$\mathcal{F}_{aba'b'}$ are the components of a non-negative 
$d_\text{A} d_\text{B} \times d_\text{A} d_\text{B}$ matrix 
$\mathcal{F}$ with the diagonal elements 1 
($\mathcal{F}_{abab}=1$), which give the representation of 
$\Lambda$. 
According to Theorem 2 and Theorem 3, the matrix 
$\mathcal{E}$ associated with the population-preserving entangling map 
$\Phi$ is inseparable, and on the other hand, the matrix 
$\mathcal{F}$ associated with the population-preserving non-entangling map
$\Lambda$ is separable. 
The above equations \eqref{eq:trEW}, \eqref{eq:w}, and \eqref{eq:trFW} give us a strategy to get an entangling-map witness. 
We first find a 
$d_\text{A} d_\text{B} \times d_\text{A}d_\text{B}$ matrix 
$\mathcal{W}$ such that 
$\text{tr}[\mathcal{E} \, \mathcal{W}] <0$ for the inseparable matrix 
$\mathcal{E}$ 
and 
$\text{tr}[\mathcal{F} \, \mathcal{W}] \geq 0$ for any separable matrix 
$\mathcal{F}$. 
Then, choosing an observable 
$\hat{W}$ satisfying Eq.\eqref{eq:w}, we get the witness $\hat{W}$ for the population-preserving entangling map 
$\Phi$.

We perform the above strategy for the case where the Hilbert space of each object is two-dimensional: 
$d_\text{A}=d_\text{B}=2$. 
Let the bases in Theorem 1 denote 
$\{ |a\,;\text{in} \rangle_\text{A} |b\,;\text{in} \rangle_\text{B} \}_{a,b=\text{L},\text{R}}$ and 
$\{ |a\,;\text{out} \rangle_\text{A} |b\,;\text{out} \rangle_\text{B} \}_{a,b=\text{L},\text{R}}$. 
We then find the following representation of a population-preserving CPTP  map 
$\Phi$ on the objects,
\begin{equation}
\Phi [\rho_\text{in}]
 =\sum_{a,b=\text{L},\text{R}} 
 \sum_{a',b'=\text{L},\text{R}} \mathcal{E}_{aba'b'} 
 \hat{M}_a \otimes  \hat{N}_b \, \rho_\text{in} \, 
 \hat{M}^{\dagger}_{a'} \otimes  \hat{N}^{\dagger}_{b'}, 
\label{eq:CPTPPP2}
\end{equation}
where 
$\hat{M}_a=|a\,;\text{out} \rangle_\text{A} \langle a\,;\text{in} |$,  
$\hat{N}_b=|b\,;\text{out} \rangle_\text{B} \langle b\,;\text{in}|$ 
and 
the coefficients 
$\mathcal{E}_{aba'b'}$ form a 
$4 \times 4$ non-negative matrix
$\mathcal{E}$ with the diagonal elements 1. 
The inseparability of such a $4 \times 4$ matrix 
$\mathcal{E}$ is completely characterized by the PPT criterion:
the partial transpose 
$\mathcal{E}^{\text{T}_\text{A}}$ with the elements 
$(\mathcal{E}^{\text{T}_\text{A}})_{aba'b'}=\mathcal{E}_{{a'bab'}}$ has a negative eigenvalue if and only if the $4 \times 4$ matrix 
$\mathcal{E}$ is inseparable.  
This fact helps us to derive an entangling-map witness. 
Let 
$\mathcal{E}^{\text{T}_\text{A}}$ have a negative eigenvalue and let
$\bm{w}=[w_\text{LL}, w_\text{LR},w_\text{RL},w_\text{RR}]^\text{T}$ be the eigenvector of the negative eigenvalue, that is,
\begin{equation}
\mathcal{E}^{\text{T}_\text{A}} \bm{w}=\nu \, \bm{w},
\label{eq:nu}
\end{equation}
where 
$\nu<0$. 
Choosing the matrix 
$\mathcal{W}$ defined around \eqref{eq:w} as
\begin{equation}
\mathcal{W}=(\bm{w}
\bm{w}^\dagger)^{\text{T}_\text{A}},
\label{eq:wwdg}
\end{equation}
we have
\begin{equation}
\text{Tr}[\hat{W} \rho ] 
=\text{tr}[\mathcal{E} \, \mathcal{W} ]
=\text{tr}[\mathcal{E} \, (\bm{w}
\bm{w}^\dagger)^{\text{T}_\text{A}} ]
=\text{tr}[\mathcal{E}^{\text{T}_\text{A}} \, \bm{w}
\bm{w}^\dagger ]
=\nu \, \text{tr}[\bm{w}
\bm{w}^\dagger]
=\nu \, 
\bm{w}^\dagger\bm{w}<0,
\label{eq:TrWrho}
\end{equation}
where Eq.\eqref{eq:trEW} was used in the first equality, and the last inequality follows by 
$\nu<0$. 
For the density operator 
$\sigma=\Lambda[\rho_\text{in}]$ with a population-preserving non-entangling map 
$\Lambda$, whose representation is specified by a separable 
$4\times 4$ matrix
$\mathcal{F}$, we obtain
\begin{equation}
\text{Tr}[\hat{W} \sigma ] 
=\text{tr}[\mathcal{F} \, \mathcal{W} ]
=\text{tr}[\mathcal{F} \, (\bm{w}
\bm{w}^\dagger)^{\text{T}_\text{A}} ]
=\text{tr}[\mathcal{F}^{\text{T}_\text{A}} \, \bm{w}
\bm{w}^\dagger ]
=\bm{w}^\dagger \mathcal{F}^{\text{T}_\text{A}}  \bm{w}
\geq 0
\label{eq:TrWsigma}
\end{equation}
where in the first equality Eq.\eqref{eq:trFW} was substituted, and the last inequality holds from the fact that the partial transpose 
$\mathcal{F}^{\text{T}_\text{A}}$ is non-negative because of the PPT criterion for the 
$4\times 4$ separable matrix 
$\mathcal{F}$.
Hence, the matrix 
$\mathcal{W}$ given in \eqref{eq:wwdg} satisfies that 
$\text{tr}[\mathcal{E} \, \mathcal{W}] <0$ for the inseparable matrix 
$\mathcal{E}$ 
and 
$\text{tr}[\mathcal{F} \, \mathcal{W}] \geq 0$ for any separable matrix 
$\mathcal{F}$. 

The next concern is whether we can get an entangling-map witness 
$\hat{W}$ from Eq.\eqref{eq:w}, which is explicitly written as 
\begin{equation}
[(\bm{w}
\bm{w}^\dagger)^{\text{T}_\text{A}}]_{a'b' ab}
=\langle a' b'\,;\text{out} | \hat{W} |a b\,;\text{out} \rangle \langle a b\,;\text{in}| \rho_\text{in} |a' b'\,;\text{in} \rangle,
\label{eq:wwdg2}
\end{equation}
where we used 
$\hat{M}^\text{A}_a=|a\,;\text{out} \rangle_\text{A} \langle a\,;\text{in}|$ and 
$\hat{N}^\text{B}_b=|b\,;\text{out} \rangle_\text{B} \langle b\,;\text{in}|$ and defined
$|a b\,; \text{in(out)} \rangle = |a\,;\text{in(out)} \rangle_\text{A} |b\,;\text{in(out)} \rangle_\text{B}$. 
For a simple case 
$\langle a b\,;\text{in}| \rho_\text{in} |a' b'\,;\text{in} \rangle \neq 0$, we obtain the following elements of the entangling-map witness 
$\hat{W}$,
\begin{equation}
\langle a' b'\,;\text{out} | \hat{W} |a b\,;\text{out} \rangle=\frac{
[(\bm{w}
\bm{w}^\dagger)^{\text{T}_\text{A}}]_{a'b' ab}}{
\langle a b\,;\text{in}| \rho_\text{in} |a' b'\,;\text{in}\rangle}.
\label{eq:Wcomp}
\end{equation}
In the next section, we will find that Eq.\eqref{eq:Wcomp} bridges the coherence of massive objects and the entangling map due to gravity. 

\section{Coherence as a source of entangling map due to gravity }
\label{sec:4}

For the setting in Sec.\ref{sec:2}, we have observed that the evolution of two gravitating objects is described by Eq.\eqref{eq:Sol2}.
The evolution is regarded as a population-preserving CPTP  map, which is characterized by the $4\times4$ matrix 
$\mathcal{E}_\text{G}$ with 
the elements 
$(\mathcal{E}_\text{G})_{aba'b'}=e^{i(\Phi_{ab}-\Phi_{a'b'})} \, (a,b,a',b'=\text{L},\text{R})$.
Explicitly, we can write down 
$\mathcal{E}_\text{G}$ and its partial transpose 
$\mathcal{E}^{\text{T}_\text{A}}_\text{G}$ with 
$(\mathcal{E}^{\text{T}_\text{A}}_\text{G})_{aba'b'}=(\mathcal{E}_\text{G})_{a'bab'}$
as 
\begin{equation}
\mathcal{E}_\text{G}=
\begin{bmatrix}
1 & e^{i(\Phi_\text{LL} -\Phi_\text{LR})} & e^{i(\Phi_\text{LL} -\Phi_\text{RL})} & e^{i(\Phi_\text{LL} -\Phi_\text{RR})} \\
e^{i(\Phi_\text{LR} -\Phi_\text{LL})} & 1 &  e^{i(\Phi_\text{LR} -\Phi_\text{RL})} & e^{i(\Phi_\text{LR} -\Phi_\text{RR})} \\
e^{i(\Phi_\text{RL} -\Phi_\text{LL})} & e^{i(\Phi_\text{RL} -\Phi_\text{LR})} & 1 & e^{i(\Phi_\text{RL} -\Phi_\text{RR})} \\
e^{i(\Phi_\text{RR} -\Phi_\text{LL})} & e^{i(\Phi_\text{RR} -\Phi_\text{LR})} &  e^{i(\Phi_\text{RR} -\Phi_\text{RL})} & 1 \\
\end{bmatrix} ,
\label{eq:EG}
\end{equation}
and 
\begin{equation}
\mathcal{E}^{\text{T}_\text{A}}_\text{G}=
\begin{bmatrix}
1 & e^{i(\Phi_\text{LL} -\Phi_\text{LR})} & e^{i(\Phi_\text{RL} -\Phi_\text{LL})} & e^{i(\Phi_\text{RL} -\Phi_\text{LR})} \\
e^{i(\Phi_\text{LR} -\Phi_\text{LL})} & 1 &  e^{i(\Phi_\text{RR} -\Phi_\text{LL})} & e^{i(\Phi_\text{RR} -\Phi_\text{LR})} \\
e^{i(\Phi_\text{LL} -\Phi_\text{RL})} & e^{i(\Phi_\text{LL} -\Phi_\text{RR})} & 1 & e^{i(\Phi_\text{RL} -\Phi_\text{RR})} \\
e^{i(\Phi_\text{LR} -\Phi_\text{RL})} & e^{i(\Phi_\text{LR} -\Phi_\text{RR})} &  e^{i(\Phi_\text{RR} -\Phi_\text{RL})} & 1 \\
\end{bmatrix}.
\label{eq:EGTA}
\end{equation}
For 
$0 \leq \Phi_\text{LR}+\Phi_\text{RL} -\Phi_\text{LL}-\Phi_\text{RR}\leq 2\pi$, the negative eigenvalue 
$\nu$ and its eigenvector 
$\bm{w}$ of 
$\mathcal{E}^{\text{T}_\text{A}}_\text{G}$ are
\begin{widetext}
\begin{equation}
\nu=-2\sin
\Big(
\frac{\Phi_\text{RL}+\Phi_\text{LR}-\Phi_\text{LL}-\Phi_\text{RR}}{2}
\Big), \quad
\bm{w}
=
\begin{bmatrix}
w_\text{LL} \\ 
w_\text{LR} \\
w_\text{RL} \\
w_\text{RR}
\end{bmatrix}
=
\frac{1}{2}
\begin{bmatrix}
-e^{-\frac{i}{2}(\Phi_\text{LR}-\Phi_\text{RL})} \\ 
ie^{-\frac{i}{2}(\Phi_\text{LL}-\Phi_\text{RR})} \\
-ie^{\frac{i}{2}(\Phi_\text{LL}-\Phi_\text{RR})} \\
e^{\frac{i}{2}(\Phi_\text{LR}-\Phi_\text{RL})}
\end{bmatrix}.
\label{eq:nuw}
\end{equation}
\end{widetext}
The matrix 
$(\bm{w} \bm{w}^\dagger)^{\text{T}_\text{A}} $ with the elements 
$[(\bm{w}
\bm{w}^\dagger)^{\text{T}_\text{A}}]_{a'b' ab}
=w_{ab'}
w^*_{ a'b}$ is 
\begin{widetext}
\begin{equation}
(\bm{w} \bm{w}^\dagger)^{\text{T}_\text{A}}
=
\frac{1}{4}
\begin{bmatrix}
1& 
ie^{-\frac{i}{2}(\Phi_\text{LL}-\Phi_\text{RR}-\Phi_\text{LR}+\Phi_\text{RL})} & 
i e^{-\frac{i}{2}(\Phi_\text{LL}-\Phi_\text{RR}+\Phi_\text{LR}-\Phi_\text{RL})} & 
-e^{-i(\Phi_\text{LL}-\Phi_\text{RR})
}\\ 
* & 1 
& -e^{-i(\Phi_\text{LR}-\Phi_\text{RL})} 
& -i e^{\frac{i}{2} (\Phi_\text{LL}-\Phi_\text{RR}+\Phi_\text{LR}-\Phi_\text{RL})} \\
* & * & 1 & -i e^{\frac{i}{2}(\Phi_\text{LL}-\Phi_\text{RR}-\Phi_\text{LR}+\Phi_\text{RL})} \\
* & * &* & 1
\end{bmatrix},
\label{eq:wwdgTA1}
\end{equation}
\end{widetext}
where 
$*$ s are determined by the Hermiticity of $(\bm{w} \bm{w}^\dagger)^{\text{T}_\text{A}}$. 
To get a simple form of a witness, we set the phases 
$\Phi_\text{LL},\Phi_\text{LR},\Phi_\text{RL}$ and 
$\Phi_\text{RR}$ to zero. 
Then, the above matrix 
$(\bm{w} \bm{w}^\dagger)^{\text{T}_\text{A}}$ is 
\begin{equation}
(\bm{w} \bm{w}^\dagger)^{\text{T}_\text{A}}
=
\frac{1}{4}
\begin{bmatrix}
1& i & i & -1\\ 
-i & 1 & -1 & -i  \\
-i & -1 & 1 & -i  \\
-1 & i& i & 1
\end{bmatrix}.
\label{eq:wwdgTA2}
\end{equation}
Assuming the initial state 
$\rho_\text{in}=\rho_\text{A} \otimes \rho_\text{B}$ given in Eqs. \eqref{eq:rhoi}, \eqref{eq:rhoA}, and \eqref{eq:rhoB}, we get the following Hermitian operator as  the candidate of entangling-map witness: 
\begin{align}
\hat{W}
&=\sum_{a',b'=\text{L},\text{R}} 
\sum_{a,b=\text{L},\text{R}} 
\langle a' b'\,;\text{out} | \hat{W} |ab\,;\text{out} \rangle 
| a' b'\,;\text{out} \rangle 
\langle ab\,;\text{out}|
\nonumber 
\\
&=
\sum_{a',b'=\text{L},\text{R}} 
\sum_{a,b=\text{L},\text{R}} 
\frac{
[(\bm{w}
\bm{w}^\dagger)^{\text{T}_\text{A}}]_{a'b' ab}}{
\langle ab\,;\text{in}| \rho_\text{in} |a'b'\,;\text{in} \rangle} | a' b'\,;\text{out} \rangle 
\langle a b\,;\text{out}|
\nonumber 
\\
&= 
\hat{\mathbb{I}}-c_1^{-1} \, \hat{Y}_\text{A} \otimes \hat{Z}_\text{B}
-c_2^{-1} \, \hat{Z}_\text{A} \otimes \hat{Y}_\text{B}
-(c_1 c_2)^{-1} \hat{X}_\text{A} \otimes \hat{X}_\text{B},
\label{eq:Wexplicit}
\end{align}
where 
$\hat{X}_\text{A}=|\text{L}\,;\text{out}\rangle_\text{A} \langle \text{R}\,;\text{out}|+|\text{R}\,;\text{out}\rangle_\text{A} \langle \text{L}\,;\text{out}|$, $\hat{Y}_\text{A}=-i\big( |\text{L}\,;\text{out} \rangle_\text{A} \langle \text{R}\,;\text{out}|-|\text{R}\,;\text{out}\rangle_\text{A} \langle \text{L}\,;\text{out}| \big)$, and 
$\hat{Z}_\text{A}= |\text{L}\,;\text{out} \rangle_\text{A} \langle \text{L}\,;\text{out}|-|\text{R}\,;\text{out} \rangle_\text{A} \langle \text{R}\,;\text{out}|$.
$\hat{X}_\text{B}$,
$\hat{Y}_\text{B}$, and 
$\hat{Z}_\text{B}$ are defined in the same manner. 
The obtained operator \eqref{eq:Wexplicit} corresponds to the entanglement witness proposed in \cite{Chevalier2020} by reinterpreting 
$\text{L}$
and 
$\text{R}$ with the spin degrees of freedoms and by setting 
$c_1=c_2=1$. 
The expectation value of 
$\hat{W}$ is
\begin{align}
\text{Tr}[\hat{W} \rho_\text{out}]
&=1
-\frac{1}{2} 
\big(
\sin[\Phi_\text{LL}-\Phi_\text{LR}]+\sin[\Phi_\text{LL}-\Phi_\text{RL}]
-\sin[\Phi_\text{RR}-\Phi_\text{RL}]-\sin[\Phi_\text{RR}-\Phi_\text{LR}] 
\big)
\nonumber 
\\
&
-\frac{1}{2} 
\big(\cos[\Phi_\text{LL}-\Phi_\text{RR}]+\cos[\Phi_\text{LR}-\Phi_\text{RL}]
\big).
\label{eq:expW}
\end{align}
Substituting the equations of 
$\Phi_\text{LL},\Phi_\text{LR}, \Phi_\text{RL}$ and 
$\Phi_\text{RR}$ given in \eqref{eq:phis} into the above equation and assuming a small 
$T$, we have 
\begin{equation}
\text{Tr}[\hat{W} \rho_\text{out}]
\approx 
-\frac{2L^2}{D^2-L^2} \frac{Gm_\text{A} m_\text{B}T}{\hbar D}<0.
\label{eq:expW2}
\end{equation}
Hence, the expectation value of 
$\hat{W}$ can be negative, and the operator works as the entangling-map witness.  

\begin{figure}[htbp]
  \centering
  \includegraphics[width=0.45\linewidth]{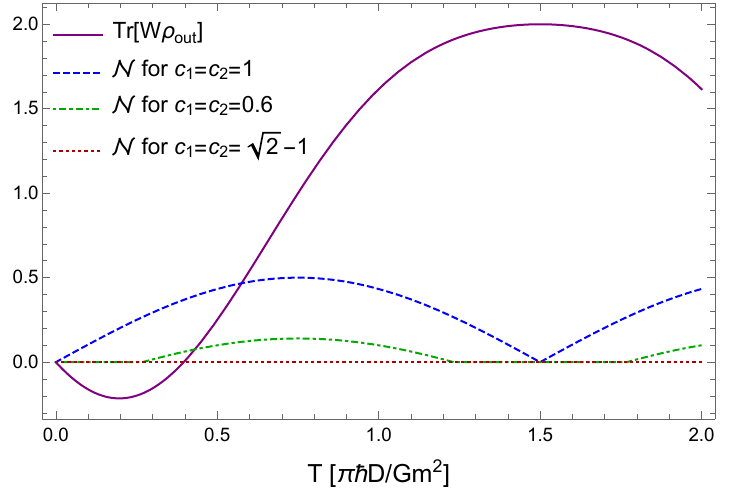}
 \mbox{\raisebox{5mm}{\includegraphics[ width=0.45\linewidth]{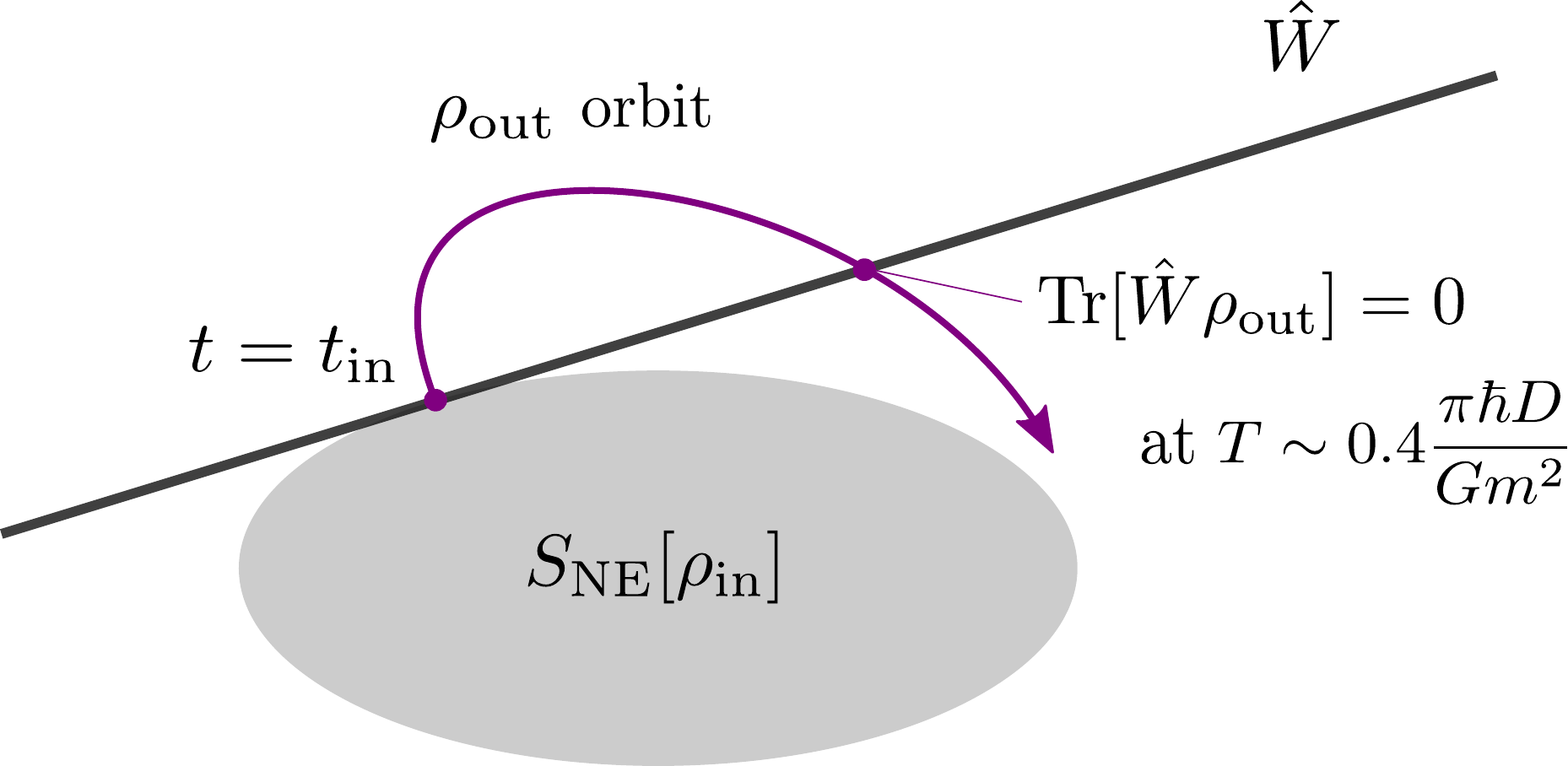}}}
  \caption{Left panel: the behavior of the negativity and the expectation value of the entangling-map witness $\hat{W}$.
  The witness becomes zero around $T \sim 0.4 \pi \hbar D/Gm^2$. 
  Right panel: the geometric picture of the entangling-map witness and the orbit of 
  the evolved state 
  $\rho_\text{out}$ in the state space. }
  \label{fig:wtnss_vs_ngtvty}
\end{figure}

The purple curve in the left panel of Fig.\ref{fig:wtnss_vs_ngtvty} shows the behavior of 
$\text{Tr}[\hat{W} \rho_\text{out}]$ given in \eqref{eq:expW} as a function of 
$T$ in the unit 
$\pi \hbar D/Gm^2 $ for fixed
$D=2L$ and 
$m_\text{A}=m_\text{B}=m$. 
The other plots present the negativity shown in Fig.\ref{fig:ngtvty}. 
The witness works well until 
$T \sim 0.4 \pi \hbar D/Gm^2$.
The right panel of Fig.\ref{fig:wtnss_vs_ngtvty} shows the geometric picture of the witness Eq.\eqref{eq:Wexplicit} and the orbit of the evolved state 
$\rho_\text{out}$ in the state space. 
Around 
$T \sim 0.4 \pi \hbar D/Gm^2$, the orbit of
$\rho_\text{out}$ crosses the hyperplane determined by 
$\hat{W}$. 
In the left panel of Fig.\ref{fig:wtnss_vs_ngtvty}, we find that the entangling map due to gravity is observed through the witness $\hat{W}$ even when the entanglement negativity is not generated for a small coherence 
(small 
$c_1$ and 
$c_2$). 

We are in the position to explain the connection between the coherence of objects and the entangling map due to gravity. 
According to \eqref{eq:Wexplicit}, 
the witness 
$\hat{W}$ is well-defined for the nonzero measures of coherence $C(\rho_\text{A})=|c_1| > 0$ and
$C(\rho_\text{B})=|c_2| > 0$. 
Then, the witness $\hat{W}$ can test the entangling map due to gravity because
$\text{Tr}[\hat{W} \rho_\text{out}]<0$ as shown in Eq.\eqref{eq:expW2}.
This means that the coherence of objects is essential to induce and probe  the entangling map due to gravity. 

This statement is very intuitive and may not be surprising. 
However, its implication is important in terms of the theorem that LOCC processes do not generate entanglement \cite{Horodecki2009}.
In the community of gravity and quantum information, as mentioned in Ref.\cite{Bose2017}, the theorem plays a key role to probe the quantum nature of gravity. 
If gravity generates entanglement, then gravity is not described by any LOCC process. 
Furthermore, when locality holds (LO is satisfied), the theorem implies that gravity can communicate quantum information content. 

The theorem on entanglement and LOCC is very useful, but it should be noted that 
the converse statement does not hold; a non-LOCC process does not always generate entanglement. 
Entanglement generation may not be essential for verifying a non-LOCC feature. 
The present result in this paper rigorously shows that not entanglement generation but the coherence of objects is crucial for probing the entangling map (non-LOCC process) due to gravity. 
 
The present witness given in Eq.(44) verifies the entangling map due to gravity. 
Intuitively, such a quantum feature seems to be caused by the quantum superposition of gravitational fields of objects with coherence.
However, the authors in Ref.\cite{Pal2021} demonstrated that two quantum probes AB can be entangled via a classical mediator M when they initially are entangled across A:BM or AM:B. 
Here, the meaning of ``classical" is that there is no quantum discord between AB and M and that M is not in a quantum superposition. 
This statement says that the entanglement between gravitating objects (probes AB) may be mediated by a classical gravitational field (a classical mediator M).

On the other hand, in Ref.\cite{Krisnanda2017}, it was shown that the entanglement generation between the probes AB signifies the nonclassicality of the mediator M in the absence of initial entanglement for A:BM or AM:B.
This says that entanglement generation can be a witness for revealing the nonclassicality of mediator under such an initial condition. 
According to this argument, it is expected that, under a suitable assumption on objects and gravitational fields, the present approach can be used to test the quantum superposition of gravitational fields.


\section{Decoherence effect on superposed particles}
\label{sec:5}

\begin{figure}[H]
  \centering
  \includegraphics[width=0.485\linewidth]{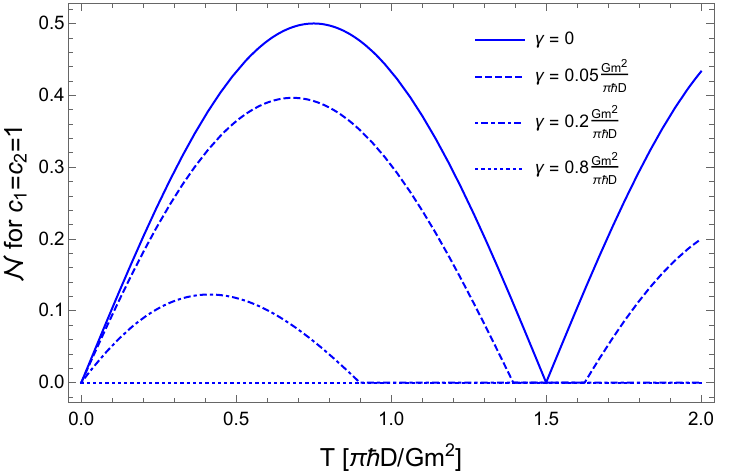}
  \includegraphics[width=0.475\linewidth]{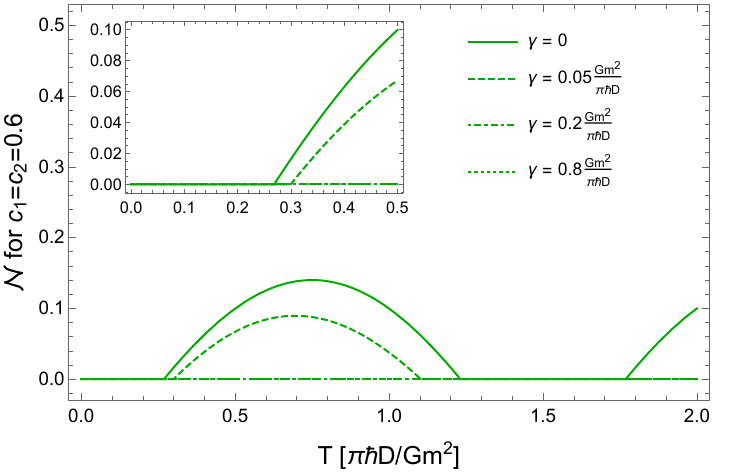}
  \caption{
The behaviors of the negativity
$\mathcal{N}$ taking the decoherence effect into account.
The left panel and the right panel present the negativity for 
$c_1=c_2=1$ and 
$c_1=c_2=0.6$, respectively.
The upper inset in the right panel shows the negativity during the time interval 
$0 \leq T \leq 0.5 \pi \hbar D/Gm^2$.
 }
  \label{fig:deco1}
\end{figure}

\begin{figure}[H]
  \centering
  \includegraphics[width=0.485\linewidth]{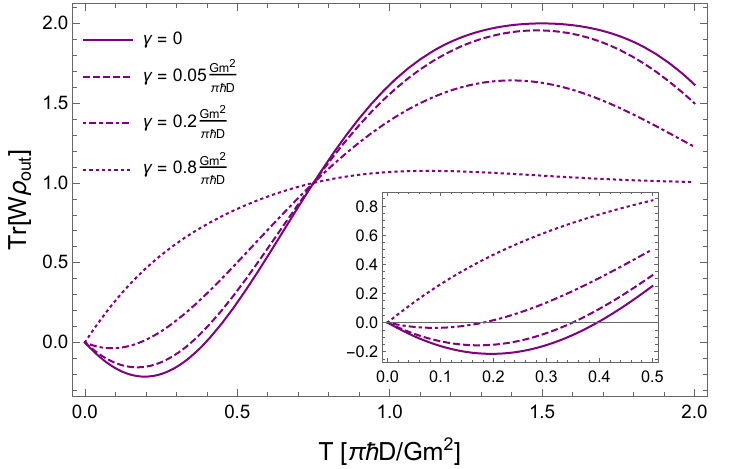}
  \caption{
The behaviors of the expectation value of the witness 
$\hat{W}$ for the various damping rates 
$\gamma$. 
The lower inset shows the behaviors for $0 \leq T \leq 0.5 \pi \hbar D/Gm^2$. As shown in Eq. \eqref{eq:expWdec}, the expectation value of the entangling-map witness obtained in the present paper does not depend on $c_1$and $c_2$.
}
  \label{fig:deco2}
\end{figure}

In this section, we discuss the decoherence suppressing the interference effect of objects.
Such suppression can be modeled by introducing the damping factor as
\begin{equation}
|\text{L}\,;\text{in} \rangle_\text{X} \langle \text{R}\,;\text{in}|
\rightarrow 
e^{-\gamma_\text{X} T} |\text{L}\,;\text{out} \rangle_\text{X} \langle \text{R}\,;\text{out}|, 
\quad 
|\text{R}\,;\text{in} \rangle_\text{X} \langle \text{L}\,;\text{in}|
\rightarrow 
e^{-\gamma_\text{X} T} |\text{R}\,;\text{out} \rangle_\text{X} \langle \text{L}\,;\text{out}|,
\label{eq:dampX}
\end{equation}
where 
$\gamma_\text{X} \, (>0)$ is the damping rate for the interference term of $\text{X}=\text{A}, \text{B}$. 
Here, we assume that the damping effect occurs during the time 
$T$ keeping the superposition (see Fig.\ref{fig:conf}). 
The damping rate depends on the length scale of superposition, the temperature for the internal degrees of freedom of objects, and other experimental parameters. 
In the present analysis, the damping rate is treated as phenomenological parameters.
Taking into account the damping factor, we find the following total evolution map 
$\tilde{\Phi}$ as
\begin{equation}
\tilde{\Phi}[\rho_\text{in}]=
\sum_{a,b=\text{L},\text{R}} 
 \sum_{a',b'=\text{L},\text{R}} (\tilde{\mathcal{E}}_\text{G})_{aba'b'} 
 \hat{M}_a \otimes  \hat{N}_b \, \rho_\text{in} \, 
 \hat{M}^{\dagger}_{a'} \otimes  \hat{N}^{\dagger}_{b'},
\label{eq:Lambda}
\end{equation}
where 
$\rho_\text{in}$ is the initial state of objects A and B, and  
$(\tilde{\mathcal{E}}_\text{G})_{aba'b'}$ are 
\begin{equation}
(\tilde{\mathcal{E}}_\text{G})_{aba'b'}=
e^{i(\Phi_{ab}-\Phi_{a'b'})} e^{-\gamma_\text{A} T  \sigma_{aa'}-\gamma_\text{B} T \sigma_{bb'}}
\label{eq:FPQ}
\end{equation}
with the gravitational phase shifts 
$\Phi_{ab}$ \eqref{eq:phis}, 
$\sigma_\text{RR}=\sigma_\text{LL}=0$ and 
$\sigma_\text{RL}=\sigma_\text{LR}=1$. 
The matrix 
$\tilde{\mathcal{E}}_\text{G}$ with the elements
$(\tilde{\mathcal{E}}_\text{G})_{aba'b'}$ is 
\begin{equation}
\tilde{\mathcal{E}}_\text{G}=
\begin{bmatrix}
1 & e^{i(\Phi_\text{LL} -\Phi_\text{LR})-\gamma_\text{B} t} & e^{i(\Phi_\text{LL} -\Phi_\text{RL})-\gamma_\text{A} t} & e^{i(\Phi_\text{LL} -\Phi_\text{RR})-(\gamma_\text{A} +\gamma_\text{B}) T} \\
e^{i(\Phi_\text{LR} -\Phi_\text{LL})-\gamma_\text{B} T} & 1 &  e^{i(\Phi_\text{LR} -\Phi_\text{RL})-(\gamma_\text{A}+\gamma_\text{B}) T} & e^{i(\Phi_\text{LR} -\Phi_\text{RR})-\gamma_\text{A} T} \\
e^{i(\Phi_\text{RL} -\Phi_\text{LL})-\gamma_\text{A} T} & e^{i(\Phi_\text{RL} -\Phi_\text{LR})-(\gamma_\text{A}+\gamma_\text{B}) T} & 1 & e^{i(\Phi_\text{RL} -\Phi_\text{RR})-\gamma_\text{B} T} \\
e^{i(\Phi_\text{RR} -\Phi_\text{LL})-(\gamma_\text{A}+\gamma_\text{B}) T} & e^{i(\Phi_\text{RR} -\Phi_\text{LR})-\gamma_\text{A} T} &  e^{i(\Phi_\text{RR} -\Phi_\text{RL})-\gamma_\text{B} T} & 1 \\
\end{bmatrix}, 
\label{eq:tildeE}
\end{equation}
and its partial transposition 
$\tilde{\mathcal{E}}_\text{G}^{\text{T}_\text{A}}$ with 
$(\tilde{\mathcal{E}}^{\text{T}_\text{A}}_\text{G})_{aba'b'} =(\tilde{\mathcal{E}}_\text{G})_{a'bab'}$ is 
\begin{equation}
\tilde{\mathcal{E}}^{\text{T}_\text{A}}_\text{G}=
\begin{bmatrix}
1 & e^{i(\Phi_\text{LL} -\Phi_\text{LR})-\gamma_\text{B} t} & e^{i(\Phi_\text{RL} -\Phi_\text{LL})-\gamma_\text{A} t} & e^{i(\Phi_\text{RL} -\Phi_\text{LR})-(\gamma_\text{A} +\gamma_\text{B}) T} \\
e^{i(\Phi_\text{LR} -\Phi_\text{LL})-\gamma_\text{B} T} & 1 &  e^{i(\Phi_\text{RR} -\Phi_\text{LL})-(\gamma_\text{A}+\gamma_\text{B}) T} & e^{i(\Phi_\text{RR} -\Phi_\text{LR})-\gamma_\text{B} T} \\
e^{i(\Phi_\text{LL} -\Phi_\text{RL})-\gamma_\text{A} T} & e^{i(\Phi_\text{LL} -\Phi_\text{RR})-(\gamma_\text{A}+\gamma_\text{B}) T} & 1 & e^{i(\Phi_\text{RL} -\Phi_\text{RR})-\gamma_\text{B} T} \\
e^{i(\Phi_\text{LR} -\Phi_\text{RL})-(\gamma_\text{A}+\gamma_\text{B}) T} & e^{i(\Phi_\text{LR} -\Phi_\text{RR})-\gamma_\text{A} T} &  e^{i(\Phi_\text{RR} -\Phi_\text{RL})-\gamma_\text{B} T} & 1 \\
\end{bmatrix}.
\label{eq:tildeETA}
\end{equation}
We examine the decoherence effect on the entanglement between the objects and the entangling map due to gravity.
We assume the same initial state   
$\rho_\text{in} = \rho_\text{A} \otimes \rho_\text{B}$ considered in Sec II. 
The entanglement negativity is evaluated as 
\begin{align}
\mathcal{N}&=\max[-\tilde{\lambda},0], 
\nonumber 
\\
\tilde{\lambda}&=\frac{1}{4}
 \Big[
 1-|c_1 ||c_2|e^{-(\gamma_\text{A}+\gamma_\text{B}) T}
\nonumber 
\\
&
 -\sqrt{(|c_1|e^{-\gamma_\text{A}T}-|c_2|e^{-\gamma_\text{B}T})^2+4|c_1||c_2|e^{-(\gamma_\text{A}+\gamma_\text{B}) T}
 \sin^2
\Big(
\frac{\Phi_\text{RL}+\Phi_\text{LR}-\Phi_\text{LL}-\Phi_\text{RR}}{2}
\Big)}
 \Big]. 
\label{eq:tildelambdadec}
\end{align}
The expectation value of the entangling-map witness 
$\hat{W}$ is 
\begin{align}
\text{Tr}[\hat{W} \rho_\text{out}]
&=1
-\frac{1}{2} 
\big(
e^{-\gamma_\text{B}T} \sin[\Phi_\text{LL}-\Phi_\text{LR}]+e^{-\gamma_\text{A}T}\sin[\Phi_\text{LL}-\Phi_\text{RL}]
-e^{-\gamma_\text{B}T}\sin[\Phi_\text{RR}-\Phi_\text{RL}]
\big)
\nonumber 
\\
&
-\frac{1}{2} 
e^{-\gamma_\text{A}T}\sin[\Phi_\text{RR}-\Phi_\text{LR}] 
-\frac{1}{2} 
e^{-(\gamma_\text{A}+\gamma_\text{B})T}\big(\cos[\Phi_\text{LL}-\Phi_\text{RR}]+\cos[\Phi_\text{LR}-\Phi_\text{RL}] 
\big).
\label{eq:expWdec}
\end{align}
We adopt the same masses
$m_\text{A}=m_\text{B}=m$, the same coherence
$c_1=c_2$ and 
the same damping rates 
$\gamma_\text{A}=\gamma_\text{B}=\gamma$.
Fig.\ref{fig:deco1} presents the behaviors of the negativity
$\mathcal{N}$ for 
$c_1=c_2=1$ (left panel) and 
$c_1=c_2=0.6$ (right panel). 
The negativity is suppressed by the decoherence effect and does not appear when
$\gamma = 0.8 Gm^2/\pi \hbar D$. 
In particular, in the right panel of Fig.\ref{fig:deco1}, for the coherence 
$c_1=c_2=0.6$, the negativity does not appear when 
$\gamma = 0.2 Gm^2/\pi \hbar D$.
On the other hand, according to the lower inset in Fig.\ref{fig:deco2}, the expectation value of 
$\hat{W}$ can be negative even when 
$\gamma = 0.2 Gm^2/\pi \hbar D$. 
Hence we can test the entangling map due to gravity for 
$\gamma = 0.2 Gm^2/\pi \hbar D$. 
We finally discuss an experimental perspective of the approach using entangling-map witness. 
As discussed in Sec.\ref{sec:2}. B, for the setting of a massive object with a spin, the parameters 
$c_1$ and 
$c_2$ are the coherence of each spin state of two objects A and B. 
To get a maximal coherence of spin
($|c_1|=|c_2|= 1$), we need a quantum control to prepare such a spin state. 
The approach in this paper works well even for a finite small coherence to verify the entangling map due to gravity. 
Hence, we do not require the procedure to prepare the state with high coherence. 
However, it is still important to manage the decoherence effect to test the entangling map due to gravity. 

\section{Conclusion}
\label{sec:6}
\vspace{-0.3cm}

We investigated the gravitational interaction between two quantum objects in terms of coherence. 
It was observed that the entanglement between the objects does not occur through their gravitational interaction, when they initially have a small amount of the $\ell_1$-norm of coherence. 
To explore the quantum nature of gravity sourced by the objects with coherence, we analyzed their time evolution adopting the notion of an entangling map. 
Identifying the evolution of the objects as population-preserving CPTP maps, we can use the theorems for those maps to proceed with the analysis. 
We obtained the entangling-map witness (the Hermitian operator) to judge whether the gravitational interaction between the objects induces an entangling map.
It was shown that the witness correctly tests the entangling map due to gravity as long as the objects initially have a finite coherence. 
We also found that the witness can capture such a feature of gravity even when the initial coherence of the objects is small and they do not get entangled. 
This means that the coherence of the gravitating objects is the source of the entangling map due to gravity. 

In the last part of this paper, we discussed a decoherence effect and an experimental advantage. 
Even in the present approach for testing the quantum nature of gravity, the decoherence effect should be suppressed. 
However, we might not need the preparation of an initial state with high coherence in the approach. 
For the approach to work well, it only requires a finite coherence of the initial state of the objects. 
We hope that this paper helps us to prove the quantum nature of gravity in future progress. 

\begin{acknowledgments}
We thank Y. Kaku, S. Kanno, S. Kukita, Y. Kuramochi, and K. Yamamoto for useful discussions and comments related to this paper. 
\end{acknowledgments}



\begin{thebibliography}{10}
\newcommand{\enquote}[1]{``#1''}
 
\bibitem{Bose2017}
S. Bose, A. Mazumdar, G. W. Morley, H. Ulbricht, M Toro$\check{\text{s}}$, M. Paternostro, A. A. Geraci, P. F. Barker, M. S. Kim, and G. Milburn, \enquote{Spin Entanglement Witness for Quantum Gravity}, 
\href{ https://doi.org/10.1103/PhysRevLett.119.240401}{Phys. Rev. Lett. \textbf{119},  240401 (2017).
}

\bibitem{Marletto2017}
C. Marletto and V. Vedral, \enquote{Gravitationally Induced Entanglement between Two Massive Particles is Sufficient Evidence of Quantum Effects in Gravity}, \href{https://doi.org/10.1103/PhysRevLett.119.240402}{Phys. Rev. Lett. \textbf{119},  240402 (2017).}

\bibitem{Nguyen2020}
H. Chau Nguyen and F. Bernards,
``Entanglement dynamics of two mesoscopic objects with gravitational interaction'',
\href{https://doi.org/10.1140/epjd/e2020-10077-8}{Eur. Phys. J. D \textbf{74}, 69 (2020).}

\bibitem{Chevalier2020}
H. Chevalier, A. J. Paige, and M. S. Kim, \enquote{Witnessing the nonclassical nature of gravity in the presence of unknown interactions},
\href{
https://doi.org/10.1103/PhysRevA.102.022428}{Phys. Rev. A \textbf{102}, 022428 (2020). }

\bibitem{vandeKamp2020}
T. W. van de Kamp, R. J. Marshman, S. Bose, and A. Mazumdar, \enquote{Quantum gravity witness via entanglement of masses: Casimir screening}, \href{
https://doi.org/10.1103/PhysRevA.102.062807}{Phys. Rev. A \textbf{102}, 062807 (2020).}

\bibitem{Miki2021}
D. Miki, A. Matsumura, and K. Yamamoto,
``Entanglement and decoherence of massive particles due to gravity'',
\href{https://doi.org/10.1103/PhysRevD.103.026017}{
Phys. Rev. D {\bf 103}, 026017 (2021).}

\bibitem{Tilly2021}
J. Tilly, R. J. Marshman, A. Mazumdar and S. Bose, \enquote{Qudits for Witnessing Quantum Gravity Induced Entanglement of Masses Under Decoherence},
\href{
https://doi.org/10.1103/PhysRevA.104.052416}{Phys. Rev. A \textbf{104}, 052416 (2021).}

\bibitem{Krisnanda2020}
T. Krisnanda, G. Y. Tham, M. Paternostro, and T. Paterek,
``Observable quantum entanglement due to gravity'', \href{https://doi.org/10.1038/s41534-020-0243-y}{
Quantum Inf. {\bf 6}, 12 (2020).}

\bibitem{Qvafort2020}
S. Qvarfort, S. Bose, and A. Serafini, ``Mesoscopic entanglement through central–potential interactions'', \href{https://iopscience.iop.org/article/10.1088/1361-6455/abbe8d/meta}{ J. Phys. B: At. Mol. Opt. Phys. \textbf{53}, 235501 (2020).}

\bibitem{Balushi2018}  A. A. Balushi, W. Cong,  and R. B. Mann,  
``Optomechanical quantum Cavendish experiment'',
\href{https://doi.org/10.1103/PhysRevA.98.043811}{Phys. Rev. A {\bf 98} 043811 (2018).}

\bibitem{Miao2020} 
H. Miao, D. Martynov, H. Yang, and A. Datta, 
``Quantum correlations of light mediated by gravity'', \href{https://doi.org/10.1103/PhysRevA.101.063804}{Phys. Rev. A {\bf 101} 063804 (2020).}

\bibitem{Matsumura2020}
A. Matsumura, K. Yamamoto,
``Gravity-induced entanglement in optomechanical systems'',
\href{https://doi.org/10.1103/PhysRevD.102.106021}{
Phys. Rev. D {\bf 102} 106021 (2020).}

\bibitem{Miki2022}
D. Miki, A. Matsumura, K. Yamamoto, 
``Non-Gaussian entanglement in gravitating masses: The role of cumulants'', 
\href{https://doi.org/10.1103/PhysRevD.105.026011}{
Phys. Rev. D \textbf{105}, 026011 (2022).}

\bibitem{Carney2021a}
D. Carney, H. Muller, and J. M. Taylor,
``Using an Atom Interferometer to Infer Gravitational Entanglement Generation'', \href{
https://doi.org/10.1103/PRXQuantum.2.030330}{Phys. Rev. X Quantum {\bf 2} 030330 (2021).}

\bibitem{Pedernales2022}
J. S. Pedernales, K. Streltsov and M. Plenio,
``Enhancing Gravitational Interaction between Quantum Systems by a Massive Mediator'', \href{https://doi.org/10.1103/PhysRevLett.128.110401}{
Phys. Rev. Lett. \textbf{128}, 110401 (2022).} 

\bibitem{Matsumura2022a}
A. Matsumura, Y. Nambu and K. Yamamoto,
``Leggett-Garg inequalities for testing quantumness of gravity'',
\href{https://doi.org/10.1103/PhysRevA.106.012214}{
Phys. Rev. A \textbf{106},012214 (2022).}

\bibitem{Bahrami2014}
M. Bahrami, A. Gro{\ss}ardt, S. Donadi and A. Bassi, ``The Schr\"{o}dinger–Newton equation and its foundations'', 
\href{https://doi.org/10.1088/1367-2630/16/11/115007}{
New J. Phys. \textbf{16}, 115007 (2014).}

\bibitem{Kafri2014}
D. Kafri, J. M. Taylor, and G. J. Milburn, ``A classical channel model for gravitational
decoherence'', 
\href{https://doi.org/10.1088/1367-2630/16/6/065020}{New J. Phys. \textbf{16}, 065020 (2014).}

\bibitem{Baumgratz2014} 
T. Baumgratz, M. Cramer, and M. B. Plenio, ``Quantifying Coherence'',
\href{https://doi.org/10.1103/PhysRevLett.113.140401}{
Phys. Rev. Lett. \textbf{113}, 140401 (2014).}

\bibitem{Harrow2003}
A. W. Harrow and M. A. Nielsen, ``Robustness of quantum gates in the presence of noise'', \href{https://doi.org/10.1103/PhysRevA.68.012308}{Phys. Rev. A \textbf{68}, 012308 (2003).}

\bibitem{Brandao2010}
F. G. S. L. Brand$\tilde{\text{a}}$o and M. B. Plenio, ``A Reversible Theory of Entanglement and its Relation to the Second Law'', \href{https://doi.org/10.1007/s00220-010-1003-1}{Commun. Math. Phys.
\textbf{295}, 829 (2010).}

\bibitem{Nielsen2002}
M. A. Nielsen and I. Chuang, \enquote{Quantum Computation and Quantum Information} (\href{https://doi.org/10.1017/CBO9780511976667}{Cambridge University Press}, Cambridge, England, 2002).

\bibitem{Matsumura2022b}
A. Matsumura, \enquote{Path-entangling operation and quantum gravitational interaction},
\href{https://doi.org/10.1103/PhysRevA.105.042425}{
Phys. Rev. A \textbf{105}, 042425 (2022).}

\bibitem{Bose2022}
S. Bose, A. Mazumdar, M. Schut, and M. Toro$\check{\text{s}}$,
\enquote{Mechanism for the quantum natured gravitons to entangle masses}, 
\href{https://doi.org/10.1103/PhysRevD.105.106028}{Phys. Rev. D \textbf{105}, 106028 (2022). }

\bibitem{Marshman2020}
R. J. Marshman, A. Mazumdar, and S. Bose,
\enquote{Locality and entanglement in table-top testing of the quantum nature of linearized gravity}, 
\href{https://doi.org/10.1103/PhysRevA.101.052110}{Phys. Rev. A \textbf{101}, 052110 (2020).}

\bibitem{Horodecki2009}
R. Horodecki, P. Horodecki, M. Horodecki, and K. Horodecki, \enquote{Quantum entanglement}, 
\href{https://doi.org/10.1103/RevModPhys.81.865}{Rev. Mod. Phys. \textbf{81}, (2009) 865.}

\bibitem{Werner1989}
R. Werner, \enquote{Quantum states with Einstein-Podolsky-Rosen correlations admitting a hidden-variable model}, \href{https://doi.org/10.1103/PhysRevA.40.4277}{Phys. Rev. A \textbf{40}, 4277 (1989).}

\bibitem{Peres1996}
A. Peres,  ``Separability Criterion for Density Matrices'', \href{https://doi.org/10.1103/PhysRevLett.77.1413}{Phys. Rev. Lett. \textbf{77}, (1996) 1413.}

\bibitem{Horodecki1996}
M. Horodecki, R. Horodecki, and P. Horodecki, ``Separability of mixed states: necessary and sufficient conditions'', 
\href{https://doi.org/10.1016/S0375-9601(96)00706-2}{
Phys. Lett. A \textbf{223}, (1996) 1-8.}

\bibitem{Vidal2002}
G. Vidal and R. F. Werner, ``Computable measure of entanglement'', \href{https://doi.org/10.1103/PhysRevA.65.032314}{Phys. Rev. A \textbf{65}, 032314 (2002).}

\bibitem{Rains1997}
E. M. Rains, ``Entanglement purification via separable superoperators'', \href{
https://doi.org/10.48550/arXiv.quant-ph/9707002}{arXiv: quant-ph/9707002(1997).}

\bibitem{Vedral1998}
V. Vedral and M. B. Plenio, ``Entanglement measures and purification procedures'', 
\href{https://doi.org/10.1103/PhysRevA.57.1619}{Phys. Rev. A \textbf{57}, 1619 (1998).}

\bibitem{Chitambar2014}
E. Chitambar, D. Leung, L. Mančinska, M. Ozols, and A. Winter, ``Everything You Always Wanted to Know About LOCC (But Were Afraid to Ask)'', 
\href{https://doi.org/10.1007/s00220-014-1953-9}{Commun. Math. Phys. \textbf{328}, 303 (2014).}

\bibitem{Cirac2001}
J. I. Cirac, W. Dür, B. Kraus, and M. Lewenstein, ``Entangling Operations and Their Implementation Using a Small Amount of Entanglement'', \href{https://doi.org/10.1103/PhysRevLett.86.544}{Phys. Rev. Lett. \textbf{86}, 544 (2001).}

\bibitem{Jamiolkowski1972}
A. Jamiolkowski, ``Linear transformations which preserve trace and positive semidefiniteness of operators", \href{https://doi.org/10.1016/0034-4877(72)90011-0}{Rep. Math. Phys. \textbf{3}, 275 (1972).}

\bibitem{Choi1975}
M.-D. Choi, ``Completely positive linear maps on complex matrices'', \href{https://doi.org/10.1016/0024-3795(75)90075-0}{Linear Algebra Appl. \textbf{10}, 285 (1975).}

\bibitem{Pal2021}
S. Pal, P. Batra, T. Krisnanda, T. Paterek, and T. S. Mahesh, ``Experimental localisation of quantum entanglement through monitored classical mediator", 	\href{https://doi.org/10.22331/q-2021-06-17-478}{Quantum \textbf{5}, 478 (2021).}

\bibitem{Krisnanda2017}
T. Krisnanda, M. Zuppardo, M. Paternostro, and T. Paterek, and T. S. Mahesh, ``Revealing Nonclassicality of Inaccessible Objects'', \href{https://doi.org/10.1103/PhysRevLett.119.120402}{Phys. Rev. Lett. \textbf{119}, 120402 (2017).}


\end{thebibliography}
\end{document}